\documentclass[preprint,%
tightenlines,%
floats,prd,eqsecnum,nobibnotes%
,nofootinbib,aps,12pt]{revtex4}
\usepackage{amsmath, amssymb}
\usepackage{xspace}
\usepackage{graphicx}
\usepackage[breaklinks=true,%
            plainpages=false,%
            bookmarks=true
           ]{hyperref}

\allowdisplaybreaks[3]
\newcommand\oldappendix{This will give an error if \backspace oldappendix
   already exists.}
\let\oldappendix\appendix
\renewcommand\appendix{\oldappendix%
   \renewcommand\theequation{\thesection.\arabic{equation}}}

\providecommand\FIGURE[2][]{\begin{figure}[#1]\begin{center}{#2}\end{center}
                       \end{figure}}

\newcounter{subfig}[figure]
\newlength{\figlen}
\renewcommand\thesubfig{(\alph{subfig})}
\newcommand\subfig[2][]{\refstepcounter{subfig}%
     \settowidth{\figlen}{{#2}}
     \begin{tabular}{c} {#2}  \\ 
     {\footnotesize \begin{minipage}{\figlen}
      \begin{center} \thesubfig\\ {#1}%
      \end{center} \end{minipage}} \end{tabular}}     

\newif\ifeprint
\eprinttrue

\newif\ifcolorfig
\ifeprint
\colorfigfalse
\else 
\colorfigtrue
\fi

\ifeprint\else
\renewcommand\S{{\text{Sec.~}}}
\fi

\newif\iftoomuchdetail
\toomuchdetailfalse
\newenvironment{detail}{\iftoomuchdetail\sf
         \setcounter{saveequation}{\value{equation}}%
         \setcounter{equation}{0}\addtocounter{detailnum}{1}%
         \renewcommand\theequation\detailtheequation%
         \fi}{
     \iftoomuchdetail%
     \ifnum\value{equation}=0\addtocounter{detailnum}{-1}\fi%
     \setcounter{equation}{\value{saveequation}}%
     \renewcommand\theequation\savetheequation%
     \fi%
     }

\newcommand\skipthis[1]{{}}

\newcommand\ct[1]{{``{#1},''}\xspace} 
\newcommand\bt[1]{{{\em #1},}\xspace} 

\providecommand\npb[3]{{{\em Nucl.\ Phys.\/} {\bf B{#1}} ({#2}) {#3}}}
\providecommand\jhep[3]{\href{http://www.iop.org/EJ/abstract/1126-6708/#2/#1/#3/}{{{\em J.\ High Energy Phys.\/} {\bf #1} ({#2}) {#3}}}}

\providecommand\lmp[3]{{{\em Lett.\ Math.\ Phys.\/} {\bf {#1}} ({#2}) {#3}}}

\providecommand\cqg[4]{\href{http://stacks.iop.org/0264-9381/#1/#4}{{{\em Class.\ Quant.\ Grav.\/} {\bf {#1}} ({#2}) {#3}}}}

\newcommand\citeprd[3]{{\href{http://link.aps.org/abstract/prd/v#1/e#3}{{\em Phys.\ Rev.\ D\/} {\bf {#1}} ({{#2}}) {#3}}}}
\newcommand\citeprl[3]{{\href{http://link.aps.org/abstract/prl/v#1/e#3}{{\em Phys.\ Rev.\ Lett.\ \/} {\bf {#1}} ({{#2}}) {#3}}}}
\providecommand\cmp[3]{{{\em Commun.\ Math.\ Phys.\/} {\bf {#1}} ({#2}) {#3}}}
\providecommand\jmp[3]{{{\em J.\ Math.\ Phys.\/} {\bf {#1}} ({#2}) {#3}}}
\providecommand\mpla[3]{{{\em Mod.\ Phys. Lett.\/} {\bf A{#1}} ({#2}) {#3}}}
\providecommand\ijmpa[3]{{{\em Int.\ J. Mod Phys.\/} {\bf A{#1}} ({#2}) {#3}}}

\providecommand\hepth[1]{\href{http://arxiv.org/abs/hep-th/#1}{{\tt arXiv:hep-th/{{#1}}}}}

\DeclareMathOperator{\re}{Re}

\newenvironment{spmatrix}{\left(\begin{smallmatrix}}{\end{smallmatrix}\right)}

\newenvironment{tinyeq}{\begingroup\tiny\ignorespaces}{%
  \endgroup\ignorespacesafterend}

\newcommand\ket[1]{\ensuremath{\lvert{#1}\rangle}}
\newcommand\bra[1]{\ensuremath{\langle{#1}\rvert}}
\newcommand\braket[2]{\ensuremath{\langle{#1}\rvert%
   \vphantom{\lvert}{#2}\rangle}}

\newcommand\anti[2]{\ensuremath{\left\{{#1},{#2}\right\}}}
\newcommand\com[2]{\ensuremath{\left[{#1},{#2}\right]}}

\newcommand\subDB[1]{\ensuremath{#1}_\text{D.B.}}

\newcommand\db[2]{\subDB{\com{#1}{#2}}}
\newcommand\dab[2]{\subDB{\anti{#1}{#2}}}

\newcommand\field[1]{{\ensuremath{\mathbb{{#1}}}}}
\newcommand\ZZ{{\field{Z}}}

\newcommand\ZR{{\field{R}}}
\newcommand\mathone{{\rlap{\kern .25em l}1}}
\newcommand\one{{\ifmmode{\text{\mathone}}\else{\mathone}\fi}}

\newcommand\transpose{{\ensuremath{\text{\sf T}}}}
\newcommand\p{\ensuremath{\partial}}
\newcommand\abs[1]{\ensuremath{\lvert{#1}\rvert}}
\newcommand\bigabs[1]{\ensuremath{\left\lvert{#1}\right\rvert}}

\newcommand\order[1]{\ensuremath{{\mathcal O}\left({#1}\right)}}

\newcommand\param{\theta} 
\newcommand\polar{\vartheta} 
\newcommand\en{\varepsilon} 
\newcommand\radwav{R} 

\newcommand\fang[3]{\braket{\vartheta,\phi}{{#1},{#2},{#3}}} 
\newcommand\normal{\ensuremath{n}}
\newcommand\vnormal{\ensuremath{\vec{\normal}}}

\newcommand\normalTag{\addtocounter{equation}{1}\tag*{\normalsize(\theequation)}}

\newif\ifLRL
\LRLfalse 
\newcommand\LRL{Laplace-Runge-Lenz vector\LRLtrue\xspace}
\newif\ifirrep
\irrepfalse 
\newcommand\irrep{\ifirrep irrep\else%
    irreducible representation (irrep)\irreptrue\fi\xspace}

\newcommand\tj{\ensuremath{\tilde{\jmath}}}

\renewcommand\theequation{\thesection.\arabic{equation}}

\makeatletter
\@addtoreset{equation}{section}
\makeatother
\begin{document}

\ifeprint
\setlength{\baselineskip}{1.2\baselineskip} 
\fi

\title{\vspace*{\fill}
Mechanics and Quantum Supermechanics of a Monopole Probe Including
a Coulomb Potential}

\author{\large Steven G. Avery}\email{avery@mps,ohio-state,edu}
\author{\large Jeremy Michelson}\email{jeremy@mps,ohio-state,edu}
\affiliation{Department of Physics 
      The Ohio State University \\
      1040 Physics Research Building \\
      191 West Woodruff Avenue \\
      Columbus, Ohio \ 43210-1117 \\ U.S.A.
      \vspace*{\fill}}

\begin{abstract}
\vspace*{\baselineskip}

A supersymmetric Lagrangian 
used to study D-particle probes
in a D6-brane background is exactly soluble.
We present an analysis of the classical and quantum mechanics
of this theory, including classical trajectories in the bosonic theory,
and the exact quantum spectrum and wavefunctions, including both bound
and unbound states.

\vspace*{\fill}

\end{abstract}
\LRLfalse 
\irrepfalse 
\preprint{\parbox[t]{10em}{\begin{flushright}
\end{flushright}}}

\maketitle

\ifeprint
\tableofcontents
\fi

\section{Introduction} \label{sec:intro}

The best understanding we have of black holes comes from string
theory.  Within string theory, those black holes which are constructed
by wrapping D-branes on cycles of a compact special holonomy manifold
have provided our deepest insights.  Nevertheless, there is still much
to be understood.  For example, while there is a good understanding of
four-dimensional black holes obtained by wrapping D2-branes on
2-cycles of a Calabi-Yau 3-fold, an analysis which includes the
addition of D6-branes wrapping the entire 3-fold has been elusive.

An additional complication has been the stability properties of
D-branes.  The appropriate basis of D-brane charges is a function of
the Calabi-Yau moduli.  Upon crossing lines of marginal stability in
moduli space, D-branes may break up into pieces corresponding to the
new stable basis.  Moreover, the attractor mechanism---or just the
fact that the moduli will vary between infinity and the black hole
horizon---can mean that lines of marginal stability are crossed in a
black hole spacetime, thus leading the microscopic description of the
black hole to be more complicated than naively thought.

This paper will not discuss these issues.  These issues have been
raised and discussed in~\cite{predenef,denef}.  One of the
consequences of the analysis in~\cite{denef} is that it becomes
interesting to understand the motion of D-particle probes in a
D6-brane background.  This is equivalently the supersymmetric quantum
mechanics of a charged particle in the background of a magnetic
monopole.  Because D0's and D6's are naively not mutually
supersymmetric---though they can form a supersymmetric bound
state~\cite{hs,wt}---the supersymmetric quantum mechanics has a
nontrivial potential, which turns out to include a Coulumb potential
(see Eq.~\eqref{L}).

Remarkably, this supersymmetric quantum mechanics is amenable to exact
analysis.
In particular, in addition to the usual conserved quantities, energy
and angular momentum%
%
, there is an additional conserved vector quantity.  The additional conserved
charges are not unlike the \LRL found in the hydrogen atom, and so we give
it the same name.

It is also interesting that attempting to understand
black holes following~\cite{denef} leads one to study
a supersymmetric quantum mechanics.  The near horizon
limit of a black hole spacetime has an AdS$_2$ factor, whose
conformal field theory dual should be a supersymmetric quantum mechanics.%
\cite{cdkktv,ms1,ms2} \
Refs.~\cite{gsx,dgmsxz}
were able to find a supersymmetric quantum mechanics by studying
the D-particles induced by D2-branes in the presence of a background
Ramond-Ramond field.
Interestingly, however, although the supersymmetric quantum mechanics
of~\cite{ms1,ms2}
was, following~\cite{cp},
``Type B'' and had fermions which are worldline spinors and target space
vectors, the supersymmetric quantum mechanics of~\cite{denef,gsx,dgmsxz}
is similar
to that of~\cite{de} in that the fermions are target space {\em spinors}.

The purpose of this paper, then, is, with the help of the \LRL, to present
the exact spectrum and wavefunctions for the theory~\eqref{L}.
Some of this was already done over 20 years ago,
for this theory or for related theories.
References include~\cite{denef,%
D'Hoker:1987ft,dv:1990,dv:1985group,dv:1985npb,dv:1986,dv:1985prl,horv:2005,%
horv:1989,horv:1987,fhr,mic,horv:1992jmp,horv:2006, ivanov}.
However, we have not seen all of the wavefunctions previously, and
many of the other results are scattered throughout the literature.
So while not all of the results in this paper are new, we have
merged our new results with the old results
in a self-contained and complete way.

As interesting as it is to have a nontrivial supersymmetric quantum mechanics
which is amenable to exact analysis, our ultimate motivation is, of course,
an understanding of black holes.  We intend for the results presented here to
be useful to that end.

It would also be interesting to understand---perhaps similar to the
relationship between the Neveu-Schwarz-Ramond superstring and the
Green-Schwarz superstring---if the supersymmetric quantum mechanics
studied here can be related to those Type B ones studied in
e.g.~\cite{ms1,ms2}.

This paper is organized as follows.
The Lagrangian of the model studied in this
paper is presented in $\S$\ref{sec:model}.
The conserved quantities and their classical and quantum algebra are
given in $\S$\ref{sec:sym}.
The classical mechanics is studied in section~$\S$\ref{sec:classical}.
The quantum spectrum for both bound and unbound states is derived
in~$\S$\ref{sec:QM:spectrum}.  The corresponding wavefunctions are presented
in~$\S$\ref{sec:states}.  Our conventions are outlined in
Appendix~\ref{sec:conventions}.  A derivation of the fermionic superpartners
to a bosonic wavefunction is given in Appendix~\ref{ap:fermWave}.
An outline of the unitary representations of SO(3,1) and its contraction,
used to study unbound and marginally bound states, is given in
Appendix~\ref{sec:SO31}.  Finally, a derivation of the $\delta$-function
normalization of the unbound bosonic states is given in
Appendix~\ref{sec:norm}.

\section{The Model} \label{sec:model}

In this paper we study the mechanics defined by the Lagrangian
\begin{equation} \label{L}
L =
\frac{m}{2}\left(\Dot{\vec{x}}^2 + D^2 + 2 i \bar{\lambda}\dot{\lambda} \right)
- \left(\frac{\kappa}{2 r} + \param\right) D
- \kappa \vec{A} \cdot \Dot{\vec{x}}
- \frac{\kappa}{2 r^3} \vec{x} \cdot \bar{\lambda} \vec{\sigma} \lambda.
\end{equation}
This and related theories have previously been studied in~\cite{%
D'Hoker:1987ft,dv:1990,dv:1985group,dv:1985npb,dv:1986,dv:1985prl,horv:2005,%
horv:1989,horv:1987,fhr,mic,horv:1992jmp,horv:2006,denef, ivanov}.  The
Lagrangian~\eqref{L} describes a particle of mass $m$ and integer
charge $\kappa$ in the background of a magnetic monopole with unit
charge.%
\footnote{Actually, only the product of the electric charge of the
test particle and the magnetic charge of the monopole appears in
the action.  This is $\kappa$.
\label{ft:kappa}}
There is also a Coulomb potential whose strength is parameterized by
the dimensionful (with units of inverse length, in $c=\hbar=1$ units)
parameter $\param$.
As the model is a three Euclidean dimensional one, the position
of the test particle, $\vec{x}=(x,y,z)$, is a three-vector whose norm
is $r = \abs{\vec{x}}$.  The fermion
$\lambda$ is a two-component spinor whose conjugate is $\bar{\lambda}$;
see Appendix~\ref{sec:conventions} for conventions.
$D$ is an auxiliary field.  The vector potential $\vec{A}$ determines
the magnetic field $\vec{B}=\vec{\nabla}\times\vec{A}$, and is the
vector potential for a unit magnetic monople,
\begin{align} \label{AB}
\vec{A} &= \begin{cases} \frac{1}{2} \bigl(1 - \frac{z}{r}\bigr) 
  \frac{x \hat{y} - y \hat{x}}{x^2+y^2}, & \frac{z}{r} > -\sin \epsilon, \\
-\frac{1}{2} \bigl(1 + \frac{z}{r}\bigr) 
  \frac{x \hat{y} - y \hat{x}}{x^2+y^2}, & \frac{z}{r} < \sin \epsilon,
\end{cases} &
\vec{B} &=\vec{\nabla} \times \vec{A} = \frac{1}{2r^3} \vec{x},
\end{align}
in terms of the unit vectors $\hat{x}$, $\hat{y}$, $\hat{z}$.
The magnetic monopole
vector potential is defined in patches which overlap in the
region \hbox{$-\sin \epsilon < \frac{z}{r} < \sin \epsilon$},
\hbox{$0 < \epsilon < \frac{\pi}{2}$}~\cite{wuyang}.
The difference between the two vector potentials in the overlap region
is pure gauge,
\begin{equation} \label{DA}
\vec{A}(z>-r \sin \epsilon) - \vec{A}(z<r \sin \epsilon)
= \vec{\nabla} \frac{1}{2} \tan^{-1} \frac{y}{x}, \qquad
-\sin \epsilon < \frac{z}{r} < \sin \epsilon
\end{equation}
Also, the choice of gauge~\eqref{AB}
gives $\vec{\nabla} \cdot \vec{A} = 0$ in both patches.

\section{Conserved Quantities} \label{sec:sym}

In terms of the spin angular momentum which is not conserved,
\begin{equation}
\vec{s} = \frac{m}{2} \bar{\lambda} \vec{\sigma} \lambda,
\end{equation}
the conserved quantities are
\begin{align} \label{H}
H &
= \frac{1}{2m} (\vec{p}+\kappa \vec{A})^2
 + \frac{m}{2} D^2
 + \frac{\kappa}{m r^3} \vec{x} \cdot \vec{s},
\\
\label{J}
\vec{J} &= \vec{x} \times (\vec{p} + \kappa \vec{A})
 + \frac{\kappa}{2r} \vec{x}
 + \vec{s},
\\
\label{K}
\vec{K} &= (\vec{p}+\kappa \vec{A}) \times \vec{J}
- i (\vec{p} + \kappa \vec{A})
+ \frac{\kappa \param}{2r} \vec{x}
+ \frac{\kappa}{r^3} (\vec{x} \cdot \vec{s}) \vec{x}
- (\frac{\kappa}{r}+\param) \vec{s}
+ (\vec{p}+\kappa \vec{A})\times \vec{s},
\\
\label{Q}
Q &= -m D \lambda 
  - i (\vec{p} + \kappa \vec{A}) \cdot \vec{\sigma} \lambda, 
\mspace{50mu}
\bar{Q} = -m D \bar{\lambda} 
  + i \bar{\lambda} \vec{\sigma} \cdot (\vec{p} + \kappa \vec{A}),
\end{align}
These quantities are respectively the Hamiltonian,
angular momentum, \LRL, and the supercharges.
The second term on the right-hand side of~\eqref{K} is a quantum correction,
also needed for hermiticity of the operator,
and should be omitted classically.
Otherwise,
these expressions are valid both classically and quantum mechanically.
$\vec{p}$
is the canonical momentum $\vec{p} = \frac{\p L}{\p \Dot{\vec{x}}}$,
and the on-shell value of $D$ is
\begin{equation}
D = \frac{1}{m} \left(\frac{\kappa}{2r} + \param\right).
\end{equation}

The first two terms of the angular momenta~\eqref{J}
can be considered to be the orbital angular momentum,
and the last term is the spin angular momentum.
Note, however, that {\em spin angular momentum and orbital angular
momentum are not separately conserved}.

The \LRL~\eqref{K} is the Noether charge associated with
the transformation
\begin{subequations}
\begin{align}
\delta \vec{x} &=  m \vec{x} \times \left(\vec{\xi} \times \Dot{\vec{x}}\right)
+ m \vec{\xi} \times \left(\vec{x}\times \Dot{\vec{x}}\right)
+ \frac{\kappa}{2r} \vec{\xi} \times \vec{x} +2 \vec{\xi} \times \vec{s},
\\
\delta \lambda &=
i m \vec{\xi} \cdot \left( \Dot{\vec{x}} \times \vec{\sigma} \right) \lambda
+ i \frac{\kappa}{2r^3}
   (\vec{\xi} \cdot \vec{x})(\vec{x} \cdot \vec{\sigma}) \lambda
- \frac{i}{2} \left(\frac{\kappa}{r} + \param\right)
  \vec{\xi} \cdot \vec{\sigma}\lambda,
\\
\delta D &=
-\frac{\kappa}{2r} \vec{\xi} \cdot \Dot{\vec{x}} 
+ \frac{\kappa}{2r^3} (\vec{\xi} \cdot \vec{x})(\vec{x} \cdot \Dot{\vec{x}})
+ \frac{\kappa}{m r^3} \vec{\xi} \cdot (\vec{x} \times \vec{s}),
\end{align}
\end{subequations}
parameterized by the vector $\vec{\xi}$.
All but the first three terms
of the \LRL~\eqref{K}
are bilinear in the fermions and so vanish in
the bosonic theory.  The fermion bilinears 
ensure that $\vec{K}$ is conserved in the supersymmetric theory.

The supercharges~\eqref{Q} are the Noether charges associated with
the transformation,
\begin{align}
\delta \vec{x} &= i \bar{\lambda} \vec{\sigma} \xi -
  i\bar{\xi} \vec{\sigma} \lambda, &
\delta D &= -\Dot{\bar{\lambda}} \xi - \bar{\xi} \dot{\lambda}, &
\delta \lambda = \Dot{\vec{x}} \cdot \vec{\sigma} \xi
 + i D \xi,
\end{align}
where the spinor $\xi$ and its conjugate $\bar{\xi}$
parameterize the supersymmetry.

The spinorial supercharges can be combined to form a third conserved vector,
\begin{equation} \label{S}
\vec{S} = \frac{1}{4H} \bar{Q} \vec{\sigma} Q.
\end{equation}
A convenient normalization has been chosen which, however, only makes the
definition~\eqref{S} well-defined away from configurations or quantum states
of zero energy.
In the nonzero energy sector of the theory, it will be
convenient to use modified angular momentum and \LRL{s}
\begin{align} \label{newJK}
\Vec{\tilde{J}} &= \vec{J} - \vec{S}, &
\Vec{\tilde{K}} &= \vec{K} + \param \vec{S}.
\end{align}

\subsection{The Classical Symmetry Algebra} \label{sec:classAlg}

As is standard,
the fermions of the theory are first order and therefore
they and their momenta are constrained.
This is also true of the auxiliary field $D$ and its canonical momentum.
It is straightforward to find the nonzero Dirac brackets
\begin{align} \label{dbs}
\db{x^i}{p_j} &= \delta^i_j, &
\db{\vec{p}}{D} &= \frac{\kappa}{2m r^3} \vec{x}, &
\dab{\lambda_\alpha}{\bar{\lambda}^\beta} &= -\frac{i}{m} \delta_\alpha^\beta.
\end{align}
As a result one finds, for the unmodified quantities,
the classical symmetry algebra
\begin{subequations} \label{classAlg}
\begin{gather}
\begin{align}
\dab{Q}{Q} &= 0 = \dab{\bar{Q}}{\bar{Q}}, & \dab{Q}{\bar{Q}} &= -2 i H \one,
\\
\db{Q}{\vec{J}} &= -\frac{i}{2} \vec{\sigma} Q, &
\db{\bar{Q}}{\vec{J}} &= \frac{i}{2} \bar{Q} \vec{\sigma}, \\
\db{Q}{\vec{K}} &= \frac{i}{2} \param \vec{\sigma} Q, &
\db{\bar{Q}}{\vec{K}} &= -\frac{i}{2} \param \bar{Q} \vec{\sigma}, \\
\db{J^i}{J^j} &= \epsilon^{ijk} J^k, &
\db{J^i}{K^j} &= \epsilon^{ijk} K^k,
\end{align} \\
\db{K^i}{K^j} = \bigl(\param^2 - 2 m H\bigr) \epsilon^{ijk} J^k
 + \frac{m}{2} \epsilon^{ijk} \bar{Q} \sigma^k Q.
\end{gather}
\end{subequations} 

Away from zero energy, we can consider the conserved vector~\eqref{S}
and the modified angular momentum and \LRL{s}~\eqref{newJK}.
These satisfy an SU(2)$\times$Spin(4), algebra%
\footnote{To be precise, whether the {\em real\/}
algebra generated by $\vec{J}$ and
$\vec{K}$ is Spin(4) or Spin(3,1) depends on whether one is considering the
sector of the theory with energies $0<E<\frac{\param^2}{2m}$ or
$E>\frac{\param^2}{2m}$.
\label{ft:spin(4)}
}
\begin{subequations} \label{classSU2Cubed}
\begin{align}
\db{S^i}{S^j} &= \epsilon^{ijk} S^k, &
\db{S^i}{\tilde{J}^j} &= 0, &
\db{S^i}{\tilde{K}^j} &= 0, \\
\db{\tilde{J}^i}{\tilde{J}^j} &= \epsilon^{ijk} \tilde{J}^k, &
\db{\tilde{J}^i}{\tilde{K}^j} &= \epsilon^{ijk} \tilde{K}^k, &
\db{\tilde{K}^i}{\tilde{K}^j} &= (\param^2 - 2 m H) \epsilon^{ijk} \tilde{J}^k.
\end{align}
\end{subequations}
The Spin(4) Casimirs are {\em classically\/} given by
\begin{align} \label{modCasimirs}
\Vec{\tilde{J}} \cdot \Vec{\tilde{K}} &= \frac{\kappa^2 \param}{4}, &
(2 m H - \param^2 ) \Vec{\tilde{J}}^2 - \Vec{\tilde{K}}^2 &= 
- \frac{\kappa^2}{2} (\param^2 - m H).
\end{align}

\subsection{The Quantum Symmetry Algebra} \label{sec:QMAlg}

Quantum mechanically, the Dirac brackets~\eqref{dbs}
become the nonzero commutators
\begin{align} \label{coms}
\com{x^i}{p_j} &= i \delta^i_j, &
\com{\vec{p}}{D} &= i \frac{\kappa}{2m r^3} \vec{x}, &
\anti{\lambda_\alpha}{\bar{\lambda}^\beta} &= \frac{1}{m} \delta_\alpha^\beta.
\end{align}
Thus the quantum symmetry algebra is essentially identical to the classical
one,
\begin{subequations} \label{QMAlg}
\begin{gather}
\begin{align}
\anti{Q}{Q} &= 0 = \anti{\bar{Q}}{\bar{Q}}, &
\anti{Q_\alpha}{\bar{Q}^\beta} &= 2 H \delta^\beta_\alpha, \\
\com{\Vec{J}}{Q} &= -\frac{1}{2} \vec{\sigma} Q, &
\com{\Vec{J}}{\bar{Q}} &= \frac{1}{2} \bar{Q} \vec{\sigma}, \\
\com{\Vec{K}}{Q} &= \frac{\param}{2} \vec{\sigma} Q, & 
\com{\vec{K}}{\bar{Q}} &= -\frac{\param}{2} \bar{Q} \vec{\sigma}, \\
\com{J^i}{J^j} &= i \epsilon^{ijk} J^k, &
\com{J^i}{K^j} &= i \epsilon^{ijk} K^k,
\end{align} \\
\label{qm:KK}
\com{K^i}{K^j} =
   i \epsilon^{ijk} 
   \left[(\param^2 - 2 m H) J^k + \tfrac{m}{2} \bar{Q} \sigma^k Q\right].
\end{gather}
\end{subequations}

Similarly, away from zero energy,%
\footnote{These expressions have been given by D'Hoker and
Vinet~\cite{D'Hoker:1987ft,dv:1990,dv:1985group,dv:1986,dv:1985prl},
who state that they require a tedius calculation.  We agree that they 
could require
a tedius calculation, for we attempted to verify them with
{\em Mathematica}$^\circledR$,
but found that 2\ifeprint GB\else\ gigabytes\fi\ of 
\ifeprint RAM \else Random Access Memory \fi
was not enough.  Nevertheless, we have at
least verified these expressions on {\em constant} test spinors.
An elegant derivation is given in~\cite{fhr}.
\label{ft:verify}}
\begin{subequations} \label{QMmodAlg}
\begin{align}
\com{S^i}{S^j} &= i \epsilon^{ijk} S^k, &
\com{\tilde{J}^i}{S^j} &= 0, &
\com{\tilde{K}^i}{S^j} &= 0, \\
\com{\Vec{S}}{Q} &= -\frac{1}{2} \vec{\sigma} Q, &
\com{\Vec{\tilde{J}}}{Q} &= 0, &
\com{\Vec{\tilde{K}}}{Q} &= 0, \\
\com{\Vec{S}}{\bar{Q}} &= \frac{1}{2} \bar{Q} \vec{\sigma}, &
\com{\Vec{\tilde{J}}}{\bar{Q}} &= 0, &
\com{\Vec{\tilde{K}}}{\bar{Q}} &= 0, \\
\com{\tilde{J}^i}{\tilde{J}^j} &= i \epsilon^{ijk} \tilde{J}^k, &
\com{\tilde{J}^i}{\tilde{K}^j} &= i \epsilon^{ijk} \tilde{K}^k, &
\com{\tilde{K}^i}{\tilde{K}^j} &=
   i \epsilon^{ijk} (\param^2 - 2 m H) \tilde{J}^k.
\end{align}
\end{subequations}
However, one of the Casimirs is quantum mechanically
modified in an important way,$^{\text{\ref{ft:verify}}}$
\begin{align} \label{QMCasimirs}
\Vec{\tilde{J}}\cdot \Vec{\tilde{K}} &= \frac{\kappa^2 \param}{4}, &
\Vec{\tilde{K}}^2 &= (2m H - \param^2) (\Vec{\tilde{J}}^2+1) 
  - \frac{\kappa^2}{2} ( m H - \param^2),
\end{align}
Also,
\begin{align} \label{SCasimir}
\vec{S}^2 &= \vec{s}^2 = -\frac{3}{4} [(m \bar{\lambda}\lambda)^2 - 
  2 m \bar{\lambda}\lambda];
\end{align}
the second term is a quantum correction.

\section{Classical Trajectories} \label{sec:classical}

In this section we consider the classical {\em bosonic\/} theory by
setting $\lambda\equiv 0$.

Poincar\'{e}~\cite{poincare} has demonstrated that trajectories of
charged particles in the presence of a magnetic monopole are always
confined to a cone whose tip lies at the monopole. The addition of a
radial potential does not effect this.  Dotting $\vec{x}$ into
Eq.~\eqref{J} yields
\begin{equation} \label{preOnCone}
\vec{J} \cdot \vec{x} = \frac{\kappa}{2} r.
\end{equation}
Spherical symmetry allows us to choose the $z$-axis to be parallel to
the angular momentum, so that $\vec{J} = \abs{\vec{J}} \hat{z}$.
Then~\eqref{preOnCone} reads
\begin{equation} \label{onCone}
\frac{z}{r} = \frac{\kappa}{2J} = \text{const}, \qquad J \equiv \abs{\vec{J}},
\end{equation}
which is the equation of a cone whose slope is
$\frac{\kappa}{\sqrt{4J^2-\kappa^2}}$.  Thus the dynamics are
constrained to the positive (negative) $z$ axis for positive
(negative) $\kappa$.  Since $z\leq r$, Eq.~\eqref{onCone} implies
\begin{equation} \label{JBound}
J \geq \frac{\abs{\kappa}}{2}.
\end{equation}

The particular radial potential in this problem admits a conserved
\LRL which, in this case, restricts the trajectories to also lie in
a plane.
Without the quantum term in Eq.~\eqref{K}, and with the fermions set to zero,
\begin{equation}
\vnormal \equiv \vec{J} -\frac{\vec{K}}{\param} = 
\left(\vec{x} + \tfrac{1}{\param} \vec{J}\right)\times(\vec{p}+\kappa\vec{A}),
\end{equation}
is conserved and orthogonal to the velocity vector.  Therefore, the
trajectories must be confined to the plane whose normal is
\vnormal
. Thus, the trajectories not only
live on the surface of a cone, they are also conic sections; however,
the plane of motion does not generally contain the origin.

To be precise, use the Casimirs~\eqref{modCasimirs} (since the modified
vectors coincide with the angular momentum and \LRL{s} when the
fermions are set to zero) to see that
\begin{equation}
\abs{\vnormal}^2 =
\bigabs{\vec{J}-\frac{\vec{K}}{\param}}^2
 = \frac{2 m H}{\param^2}\left(\vec{J}^2
 - \frac{\kappa^2}{4}\right).
\end{equation}
Notice that this is nonzero except for orbits with minimal angular momentum.
(Orbits with $J=\frac{\kappa}{2}$ lie entirely on the $z$-axis,
by~\eqref{onCone}.)
When $\vnormal$ is nonzero,
the orthogonal distance from the plane of motion to the origin is
\begin{equation} \label{orthogDist}
z' \equiv \vec{x} \cdot \frac{\vnormal}{\abs{\vnormal}} =
\sqrt{\frac{\vec{J}^2-\frac{\kappa^2}{4}}{2mE}},
\end{equation}
where $E$ is the energy.

We can complete the definition of the primed coordinates $(x',y',z')$.
First complete the definition of the {\em un\/}primed coordinates
by choosing
the $y$ direction (so far only the $z$ direction was chosen)
so that the normal vector $\vnormal$ has no $y$ component.
Then choose $x'$ to be only a rotation of the $x$ and $z$ directions,
i.e.,
\begin{align} \label{primedCoords}
\begin{aligned}
\hat{x}' &= \frac{\normal_z}{\abs{\vnormal}} \hat{x}
  - \frac{\normal_x}{\abs{\vnormal}} \hat{z}, \\
\hat{y}' &= \hat{y}, \\
\hat{z}' &= \frac{\vnormal}{\abs{\vnormal}},
\end{aligned} &&
\begin{aligned}
\hat{x} &= \frac{\normal_z}{\abs{\vnormal}} \hat{x}'
  + \frac{\normal_x}{\abs{\vnormal}} \hat{z}', \\
\hat{y} &= \hat{y}', \\
\hat{z} &= -\frac{\normal_x}{\abs{\vnormal}} \hat{x}'
  + \frac{\normal_z}{\abs{\vnormal}} \hat{z}'
\end{aligned}
\end{align}
Since $\vec{J} \equiv J \hat{z}$,
\begin{align}
\normal_z &= \frac{\vec{J}}{J}\cdot \vnormal
 = J - \frac{\kappa^2}{4J},
&
\normal_x &= \sqrt{\vnormal^2-\normal_z^2}
 = \sqrt{\left(\frac{2mE}{\param^2} - 1 + \frac{\kappa^2}{4J^2}\right)
\left(J^2 - \frac{\kappa^2}{4}\right)}.
\end{align}
It is shown below (see remark~\ref{classlist:bound} on
p.~\pageref{classlist:bound}) that the first
factor under the square root is indeed positive, as required for consistency.

Using $r=\sqrt{x^{\prime2}+y^{\prime2}+z^{\prime2}}$
and $z = (x' \hat{x}' + y' \hat{y}' + z' \hat{z'}) \cdot \hat{z}$
where $z'$ is the constant~\eqref{orthogDist},
the equation for the cone~\eqref{onCone} reads,
\begin{subequations}
\begin{equation}
(1-\epsilon^2)(x'-x'_0)^2 - 2 \epsilon r_0 (x'-x_0') + y'^2 = r_0^2,
\end{equation}
where
\begin{align}
\epsilon &= \sqrt{\frac{2J^2}{\kappa^2 m E}
   \left(2mE - \param^2 + \frac{\kappa^2\param^2}{4 J^2}\right)}, &
x'_0 &= -\,
   \frac{\sqrt{J^2 - 
                \frac{\param^2}{2mE}\left(J^2-\frac{\kappa^2}{4}\right)}}
{         \abs{\param}-\sqrt{2mE}}, &
r_0 &= \frac{4J^2-\kappa^2}{2 \abs{\kappa}\sqrt{2mE}}.
\end{align}
\end{subequations}
This is easily
recognized as a conic section in the $(x',y')$-plane.
$\epsilon$ is
the eccentricity of the orbit, $r_0$ is
the semi-latus rectum of the conic section, and
the foci are
offset along the $x'$ axis, with one at $x'=x'_0$ and the other
at $x'=x'_0 + \frac{2\epsilon}{1-\epsilon^2}r_0 =
-\frac{\sqrt{J^2 - 
        \frac{\param^2}{2mE}\left(J^2-\frac{\kappa^2}{4}\right)}}
        {\abs{\param}+\sqrt{2mE}}$.
In the unprimed coordinate system, the foci are located at
\begin{equation}
\vec{x} = \frac{\sqrt{1 - \frac{\param^2}{2m E} + 
  \frac{\kappa^2\param^2}{8mE}}\sqrt{J^2-\frac{\kappa^2}{4}}}{
\sqrt{2m E} \pm \abs{\param}} \hat{x}
+ \frac{1}{J\sqrt{2m E}}\left(\pm J^2 - \frac{\frac{\kappa^2}{4} \abs{\param}}{
   \sqrt{2mE} \pm \abs{\param}}\right)\hat{z}.
\end{equation}
A bound and an unbound orbit are shown in Fig.~\ref{fig:orbits}.

\FIGURE[t]{
  \begin{tabular}{cc}
\ifcolorfig
    \subfig{\includegraphics[width=2in,clip=true]{conePlaneBoundOrbit.eps}}
    \label{fig:orbits:fullBound} &
    \subfig{\includegraphics[width=2in,clip=true]{boundOrbit.eps}}
    \label{fig:orbits:bound} \\
    \subfig{\includegraphics[width=2in,clip=true]{conePlaneUnboundOrbit.eps}}
    \label{fig:orbits:fullUnbound} &
    \subfig{\includegraphics[width=2in,clip=true]{unboundOrbit.eps}}
    \label{fig:orbits:unbound}
\else
    \subfig{\includegraphics[width=2in,clip=true]{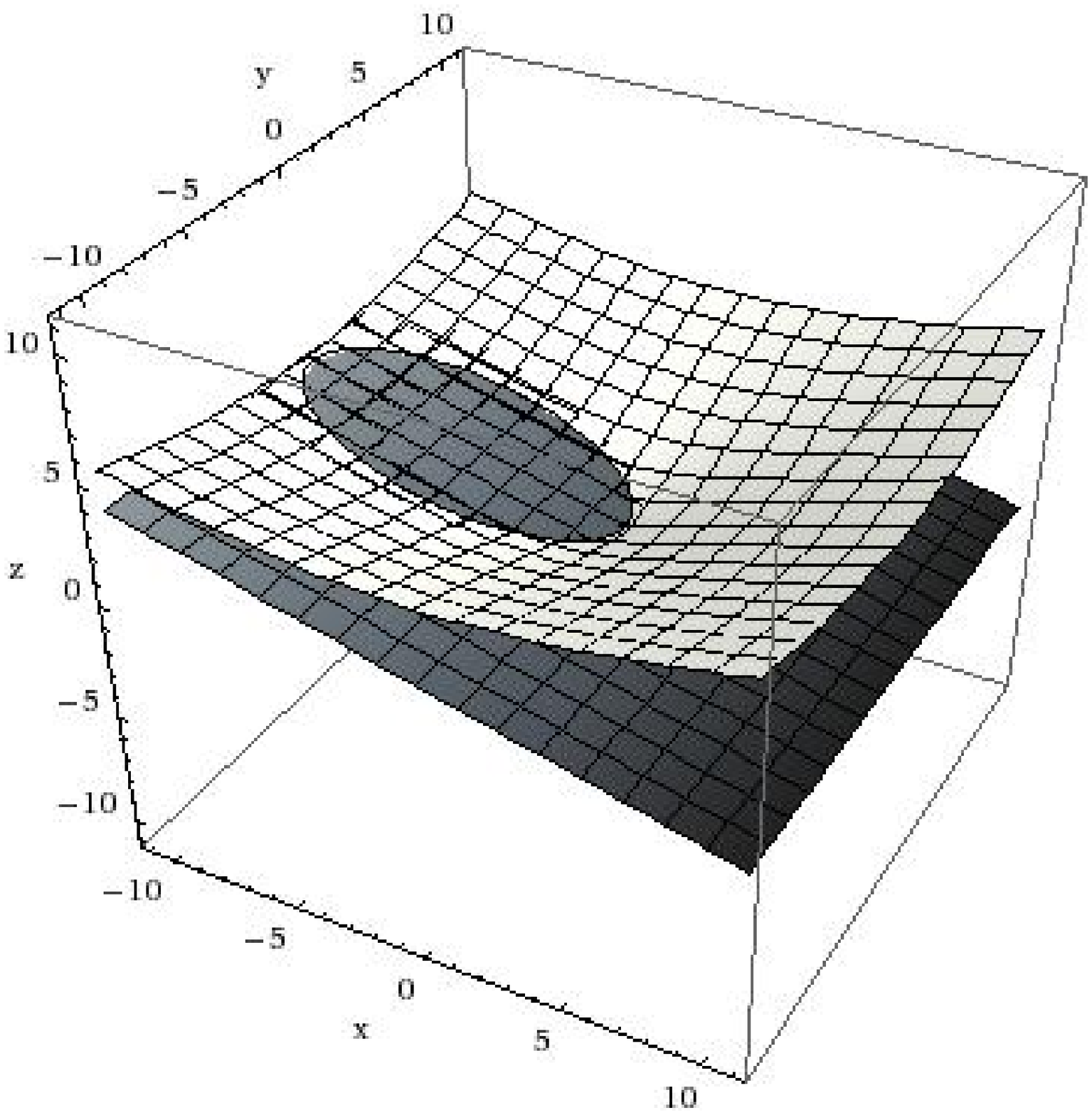}}
    \label{fig:orbits:fullBound} &
    \subfig{\includegraphics[width=2in,clip=true]{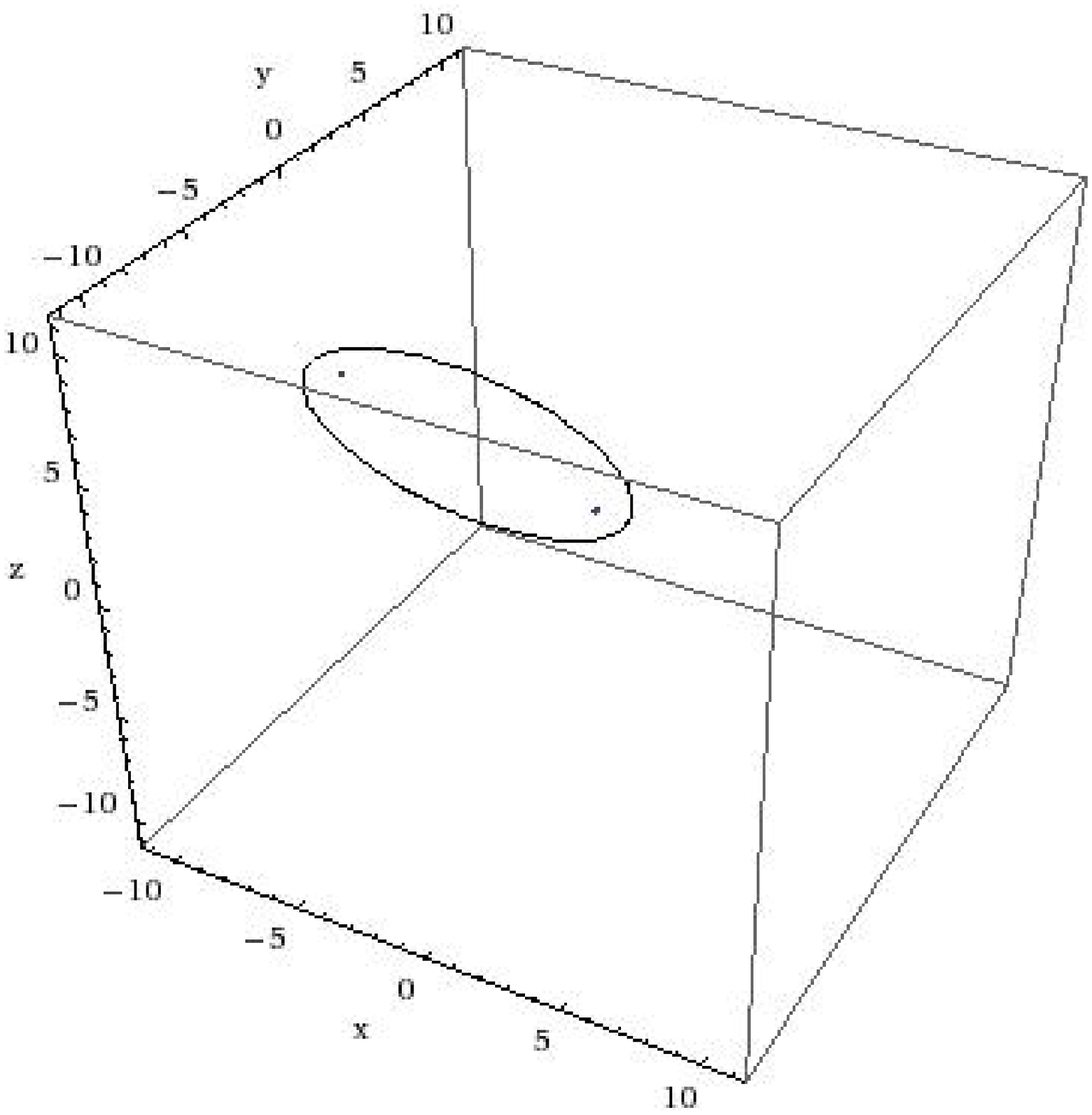}}
    \label{fig:orbits:bound} \\
    \subfig{\includegraphics[width=2in,clip=true]{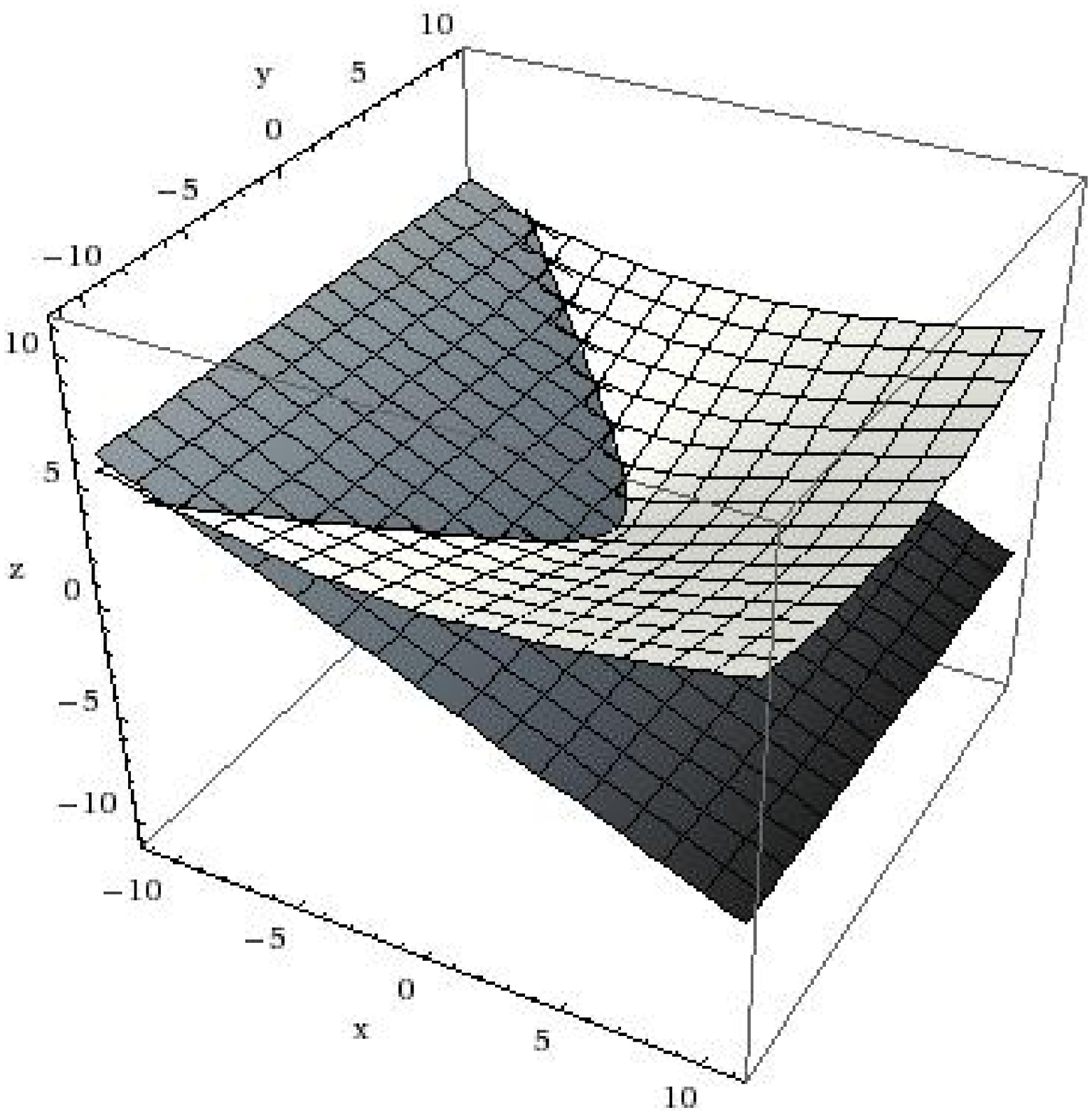}}
    \label{fig:orbits:fullUnbound} &
    \subfig{\includegraphics[width=2in,clip=true]{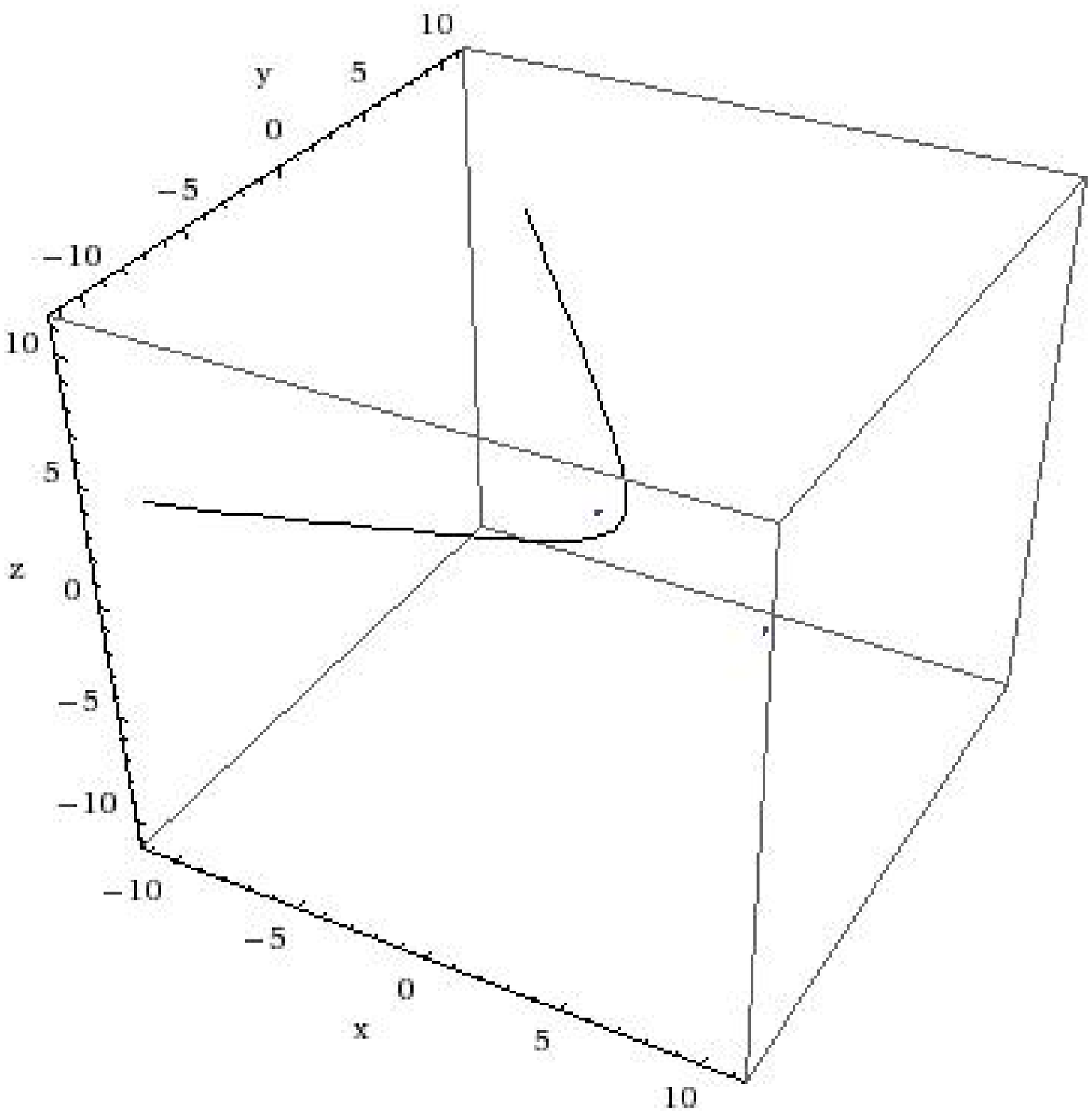}}
    \label{fig:orbits:unbound}
\fi
  \end{tabular}
\caption{Two representative orbits, one bound and one unbound,
with $J=\frac{3}{2}$,
$\kappa=1$ and units in which $\param=-2$.
\ref{fig:orbits:fullBound} and~\ref{fig:orbits:bound}
show a bound orbit with $m E=1.92089$ with and without explicitly exhibiting
the plane and the cone on which the orbit lies.  In particular, the foci are
visible in~\ref{fig:orbits:bound}.  \ref{fig:orbits:fullUnbound}
and~\ref{fig:orbits:unbound} are similar with $m E = 2.2$.
\label{fig:orbits}}
}

To find the explicit time-dependence of the particle on its orbit, we
return to the unprimed coordinates.
Upon using cylindrical coordinates
$(\rho=\sqrt{x^2+y^2},\phi=\tan^{-1}\frac{y}{x},z)$,
Eq.~\eqref{onCone} yields
\begin{equation} \label{rho}
\rho(t) = \frac{\sqrt{4J^2-\kappa^2}}{\kappa} \; z(t)
 = \sqrt{\frac{4J^2}{\kappa^2} - 1} \; \abs{z(t)}.
\end{equation}
Thus, knowledge of the trajectory reduces to finding
the two coordinates $z(t)$ and $\phi(t)$.  They are found by using
the conserved energy and angular momentum to yield first order
differential equations.

Recalling the choice of $z$-axis parallel to $\vec{J}$, the $z$-component
of angular momentum is (since $m \Dot{\vec{x}} = \vec{p}+\kappa \vec{A}$)
\begin{equation} \label{Jeom}
J = m \rho^2 \dot{\phi} + \frac{\kappa z}{2r}
\qquad \Leftrightarrow \qquad
z^2 \dot{\phi} = \frac{\kappa^2}{4m J}.
\end{equation}
Combining~\eqref{Jeom} with conservation of energy, $E$, then yields
\begin{equation} \label{Eeom}
\dot{z}^2 = \frac{\kappa^2}{4m^2 J^2} \left[ (2m E  - \param^2)
 - \frac{\kappa^2}{4 z^2}
 - \frac{\kappa^2\param}{2 Jz} \right],
\end{equation}
which is also valid for $J=\frac{\kappa}{2}$.
From~\eqref{Eeom} we can see a qualitative difference between
$E > \frac{\param^2}{2m}$ and $E < \frac{\param^2}{2m}$, for
the turning points of the motion are located at
\begin{equation} \label{zmin}
z_{\begin{smallmatrix}\text{min}\\\text{max}\end{smallmatrix}}
= \left[-\frac{\param}{J} \pm 
\frac{2}{\kappa} \sqrt{\frac{\kappa^2 \param^2}{4 J^2}
  + (2m E - \param^2)} \right]^{-1}.
\end{equation}
Because $z$ and $\kappa$ have the same sign,
\begin{enumerate}
\item \label{classlist:sameSign} if $\kappa\param > 0$ then
only the choice of plus sign
in~\eqref{zmin} is valid, and even then only if $2m E > \param^2$.
From~\eqref{Eeom}, $\abs{z}=\infty$ is allowed ($\dot{z}^2\geq 0$)
in this case, so
the single turning point gives the minimal value of $z$, and
the orbits are unbound.

\item \label{classlist:bound}
if $\kappa \param < 0$ and $2mE \leq \param^2$ then both turning
points are allowed, and the orbits are bound.  Reality of the orbit
requires $2mE\geq \param^2 - \frac{\kappa^2 \param^2}{4 J^2} \geq 0$,
upon using~\eqref{JBound} for the rightmost inequality.  The rightmost
bound also follows because the Hamiltonian is a sum of squares.

\item \label{classlist:unbound}
if $\kappa \param < 0$ and $2mE > \param^2$ then only the choice of
plus sign in~\eqref{zmin} is valid, again corresponding to unbound orbits.

\end{enumerate}
For unbound orbits with $2mE=\param^2$, $z_\text{min}=\infty$, and so
the particle is stuck at infinity.

In any case, the solutions to~\eqref{Jeom} and~\eqref{Eeom} are
\begin{equation} \label{phisoln}
\phi - \phi_0 =
\cos^{-1} \frac{\abs{\kappa} (J+\param z)}{\sqrt{4 J^2(2m E - \param^2)
   + \kappa^2 \param^2} \; z},
\end{equation}
and
\begin{subequations}
\begin{multline} \label{tsoln}
t - t_0 = \frac{2 m J}{\abs{\kappa} (2m E - \param^2)}
\left[ \sqrt{(2mE-\param^2)z^2 - \frac{\kappa^2\param}{2 J} z
  - \frac{\kappa^2}{4}}
\right. \\ \left.
+ \frac{\kappa^2\param}{4 J \sqrt{2m E - \param^2}}
\cosh^{-1} \frac{4 J (2 m E - \param^2) z - \kappa^2 \param}{
   \sqrt{4 J^2 \kappa^2 (2 m E - \param^2) + \kappa^4 \param^2}}
\right], \qquad 2m E \neq \param^2,
\end{multline}
or
\begin{equation} \label{tsolnThreshhold}
t - t_0 = \frac{8 J^2}{3\abs{\kappa}^3\param^2}
  \sqrt{\frac{\param}{2J} z - \frac{1}{4}}
  \left[\frac{\kappa^2 \param}{2J z - \frac{\kappa^2}{4} + 3}\right], \qquad
2 m E = \param^2
\end{equation}
\end{subequations}
For $2m E < \param^2$, one can replace
$\frac{1}{\sqrt{2mE-\param^2}} \cosh^{-1}$ with
$\frac{1}{\sqrt{\param^2-2mE}} \cos^{-1}$, because
Eq.~\eqref{Eeom} and the inequality in remark~\ref{classlist:bound}
above ensure that the arguments of the other
square roots are always positive.  Moreover, the
upper bound from Eq.~\eqref{zmin} implies that the
argument
of the $\cosh^{-1}$ has magnitude less than one and so should be written
as $i \cos^{-1}$.  (For the unbound orbits, the lower bound from
Eq.~\eqref{zmin}
implies that the argument of the $\cosh^{-1}$ is never less than one.)

\section{The Quantum Spectrum} \label{sec:QM:spectrum}

The quantum spectrum has been worked out by D'Hoker and
Vinet~\cite{D'Hoker:1987ft,dv:1990,dv:1985group,dv:1986,dv:1985prl,dv:1985npb};
we
repeat the analysis here for notation and completeness.
The method is algebraic; the energy spectrum and degeneracy
can be extracted from the algebra of the (modified) angular momentum and
\LRL{s}.

The Hilbert space is a product of a bosonic Hilbert space and the Hilbert
space of the $\lambda$'s.  The latter is defined by choosing the $\lambda$'s
to be annihilation operators and the $\bar{\lambda}$'s to be creation
operators, as per
the anticommutation relation~\eqref{coms}.  Using $\ket{00}$ as the
``vacuum'' state
annihilated by both $\lambda_\alpha$'s, the action of
$\bar{\lambda}$ creates the other three states
\begin{align} \label{Fstates}
\ket{10} &= \sqrt{m} \bar{\lambda}^1 \ket{00}, &
\ket{01} &= \sqrt{m} \bar{\lambda}^2 \ket{00}, &
\ket{11} &= m \bar{\lambda}^1 \bar{\lambda}^2 \ket{00}
  = -m \bar{\lambda}^2 \bar{\lambda}^1 \ket{00}.
\end{align}
Note the choice of sign for $\ket{11}$ and the normalization
which ensures
\begin{equation} \label{FNorm}
\braket{ab}{a'b'} = \delta_{aa'} \delta_{bb'},
\end{equation}
and
\begin{equation} \label{sOnF}
\begin{gathered}
\vec{s} \ket{00} = 0 = \vec{s} \ket{11}, \\
\begin{aligned}
s^3 \ket{10} &= \frac{1}{2} \ket{10}, & s^+ \ket{10} &= 0, &
s^- \ket{10} &= \ket{01}, \\
s^3 \ket{01} &= -\frac{1}{2} \ket{01}, & s^+ \ket{01} &= \ket{10}, &
s^- \ket{01} &= 0.
\end{aligned}
\end{gathered}
\end{equation}
The fermion number is $m\bar{\lambda}\lambda$.
Although $\vec{s}$ is not conserved by the Hamiltonian,
fermion number is a good quantum number.

As was foreshadowed by the introduction of the operators
$\vec{S}, \Vec{\Tilde{J}}$ and $\Vec{\Tilde{K}}$, the analysis will be
separated into zero energy and nonzero energy states.  Because
there is no central charge in the superalgebra~\eqref{QMAlg}, the
supersymmetric states precisely coincide with the states of zero energy
and are annihilated by {\em all\/} the supercharges.  The states of
nonzero energy are never annihilated by a {\em real} supercharge
(e.g.\ $Q_1 + \bar{Q}^1$), but
may be annihilated by a {\em complex} supercharge.

\subsection{Supersymmetric Ground States} \label{sec:susyspectrum}

Zero energy states---if they
exist---are annihilated by $Q$ and by $\bar{Q}$.
Eq.~\eqref{qm:KK} shows that
on such states,
\begin{equation} \label{N+-}
\vec{N}_\pm \equiv \frac{1}{2}(\vec{J} \pm \frac{1}{\param} \vec{K}),
\end{equation}
are a pair of canonically normalized commuting SU(2) generators.
One can show that
\begin{equation} \label{JK}
\vec{J}\cdot \vec{K} = \frac{\param \kappa^2}{4}
- m (\frac{\kappa}{2r} + \param) \bar{\lambda}\lambda
+ m^2 (\frac{\kappa}{2r} + \frac{3 \param}{4}) (\bar{\lambda}\lambda)^2
+ \frac{m}{2} \vec{x} \cdot [(\vec{p}+\kappa \vec{A}) \times
     (\bar{\lambda} \vec{\sigma} Q)]
+ \frac{m}{2} \bar{\lambda} Q,
\end{equation}
and
\begin{multline} \label{K2}
\vec{K}^2 
= (2 m H - \param^2) (\vec{J}^2+1)
- \frac{\kappa^2}{2}(m H - \param^2)
+ \frac{m^2}{2}
     [(\bar{\lambda}\lambda)^2 - \frac{2}{m} \bar{\lambda}\lambda]
     [\param(\frac{2 \kappa}{r} + \param)  -5 m H]
+ m \param (\frac{\kappa}{r} + \param) \bar{\lambda}\lambda
\\
+ m \param [\vec{x} \times (\vec{p} + \kappa \vec{A})]\cdot
    \bar{\lambda} \vec{\sigma} Q
+ 2 m^2 [\vec{x} \times (\vec{p} + \kappa \vec{A})]\cdot
    \bar{\lambda} \vec{\sigma} \lambda H
- \frac{m^2 \kappa}{r} \vec{x} \cdot \bar{\lambda} \vec{\sigma} \lambda H
+ m \param \bar{\lambda} Q.
\end{multline}
Thus, {\em on zero energy states\/}
\begin{align} \label{JK,E=0}
\vec{J}\cdot \vec{K} &= \frac{\param \kappa^2}{4}
- \frac{1}{4} m^2 \param (\bar{\lambda}\lambda)^2
+ m(\frac{\kappa}{2r}+\param)[(\bar{\lambda}\lambda)^2
  - \frac{1}{m} \bar{\lambda}{\lambda}], \\ \label{K2,E=0}
\vec{K}^2 &= -\param^2 (\vec{J}^2-\frac{\kappa^2}{2}+1)
+ \frac{m \param^2}{2} \bar{\lambda}\lambda
+ \frac{m^2 \param}{2}(\frac{2 \kappa}{r} + \param)
      [(\bar{\lambda}\lambda)^2 - \frac{1}{m} \bar{\lambda}\lambda].
\end{align}
Except on states with fermion number $1$, these have $r$ dependence,
which is incompatible with their conservation.  In particular,
the Casimirs
\begin{equation}
\vec{N}_\pm^2 = (1\pm 1) \frac{\kappa^2}{8} - \frac{1}{4} + \frac{1}{8}(1\mp 1)
  + f_\pm(r) [(\bar{\lambda}\lambda)^2 - \frac{1}{m} \bar{\lambda}\lambda],
\end{equation}
in terms of some functions $f_\pm(r)$ whose precise form we will not need,
are not constant except on states with fermion number $1$.
Therefore, zero energy states must be fermionic.

On the fermionic ground states, then,
\begin{align} \label{NCasimir}
\vec{N}_+^2 &= \frac{1}{4}(\kappa +1)(\kappa-1), &
\vec{N}_-^2 &= 0,
\end{align}
corresponding to a singlet under $\vec{N}_-$
and one multiplet of degeneracy
$\abs{\kappa}$ under $\vec{N}_+$.%
\footnote{To be precise, this argument so far
only ensures that the ground states appear in multiples of
$\abs{\kappa}$.  However, a ``highest weight'' state will be an
eigenstate of $N^3_\pm$---and therefore of $J^3$---and be annihilated
by $N^+_\pm$---and therefore $J^+$ and $K^+$.  This determines the
angular part of the state in the multiplet and reduces the
determination of the highest weight state to a first order radial
differential equation which has {\em at most\/} one {\em
normalizable\/} solution.  Explicitly solving the wave equation yields
one multiplet of ground states when $\kappa\param<0$ but no zero
energy states if $\kappa \param>0$.
\label{ft:deg}}
Since $\vec{J} = \vec{N}_+ + \vec{N}_-$, the ground states thus form
a multiplet whose angular momentum is $\frac{\abs{\kappa}-1}{2}$.

\subsection{\skipthis{SU(2)$^3$ Basis of }Nonsupersymmetric Bound States}
\label{sec:boundspectrum}

To find the spectrum of the remaining bound states,
first define the operators
\begin{equation} \label{defmodN}
\Vec{\tilde{N}}_{\pm} = \frac{1}{2}\left(\Vec{\tilde{J}} \pm
  \frac{1}{\sqrt{\param^2-2mH}} \Vec{\tilde{K}}\right).
\end{equation}
Notice that these operators are Hermitian only on states with
$2mE < \param^2$, where $E$ is the energy.
The operators~\eqref{defmodN} are constructed to obey the SU(2)$^2$ algebra
\begin{equation} \label{modNAlg}
\com{\tilde{N}_\pm^i}{\tilde{N}_{\pm'}^j} = \delta_{\pm\pm'} i \epsilon^{ijk}
  \tilde{N}_\pm^k.
\end{equation}
They also commute with the third SU(2), $\vec{S}$.  These three SU(2)'s
are symmetries, and so we can label the states by their quantum numbers,
though we will see that they are not completely independent.

Upon using Eq.~\eqref{modCasimirs}, the Casimirs of $\Vec{\tilde{N}}_\pm$ are
\begin{equation} \label{modNCasimir}
\Vec{\tilde{N}}_\pm^2 = -\frac{1}{4}
  + \frac{\kappa^2}{16}
    \left(-\frac{\param}{\sqrt{\param^2 - 2 m H}} \mp 1\right)^2.
\end{equation}
But of course, on a given SU(2)$^2$ state,
\hbox{$\Vec{\tilde{N}}_\pm^2 = n_\pm(n_\pm+1)
 = (n_\pm+\frac{1}{2})^2-\frac{1}{4}$},
with $n_\pm$ half-integer.
Thus,
\begin{equation}
n_\pm + \frac{1}{2} =
\frac{\abs{\kappa}}{4} \left(-\frac{\param}{\sqrt{\param^2 - 2 m E}}
   \mp 1\right).
\end{equation}
Therefore,
\begin{equation}
n_- - n_+ = \frac{\abs{\kappa}}{2}.
\end{equation}
In particular, $n_-$ is {\em larger\/} than $n_+$.
Also,
\begin{equation}
n_+ + n_- + 1 = 2n_- -\frac{\abs{\kappa}}{2} + 1
= -\frac{\abs{\kappa}\param}{2\sqrt{\param^2 - 2mE}},
\end{equation}
from which we read off the energy
\begin{equation} \label{2mE}
2mE
= \frac{4 n \param^2 (n - \abs{\kappa})}{(2n-\abs{\kappa})^2},
\qquad n \equiv 2n_- + 1.
\end{equation}
Eq.~\eqref{2mE} defines the positive, integer quantum number $n$.  Since
$n_- = n_+ + \frac{\abs{\kappa}}{2}$ and $n_+$ is positive semi-definite,
\begin{equation}
n = \abs{\kappa}+1, \abs{\kappa}+2, \dotsc.
\end{equation}

So far only $\vec{N}_\pm$, has been used and not $\vec{S}$.
Because $\vec{S}$ commutes with $\vec{N}_\pm$ and the Casimir~\eqref{SCasimir}
only depends on the fermion number, fermion number is a good simultaneous
quantum number.
In particular, consider the nonzero bound states with fermion number zero.
These are denoted by
\begin{subequations} \label{nbasis}
\begin{equation} \label{bound0}
\ket{n;00;n^-_3,n^+_3}, \qquad 
\begin{aligned}
n^-_3 &= -\tfrac{n-1}{2},-\tfrac{n-1}{2}+1,\dotsc,
  \tfrac{n-1}{2}-1,\tfrac{n-1}{2}, \\
n^+_3 &= -\tfrac{n-\abs{\kappa}-1}{2}, -\tfrac{n-\abs{\kappa}-1}{2}+1, \dotsc,
  \tfrac{n-\abs{\kappa}-1}{2}-1, \tfrac{n-\abs{\kappa}-1}{2}.
\end{aligned}
\end{equation}
These states are annihilated by both $Q$s.
The $\bar{Q}$s generate three more sets of states,
\begin{equation}
\begin{gathered}
\begin{aligned} \label{boundrest}
\ket{n;10;n^-_3,n^+_3} &\equiv \frac{1}{\sqrt{2m E}} \bar{Q}^1
\ket{n;00;n^-_3,n^+_3}, &
\ket{n;01;n^-_3,n^+_3} &\equiv \frac{1}{\sqrt{2m E}} \bar{Q}^2
\ket{n;00;n^-_3,n^+_3},
\end{aligned} \\
\ket{n;11;n^-_3,n^+_3} \equiv \frac{1}{2m E} \bar{Q}^1 \bar{Q}^2
\ket{n;00;n^-_3,n^+_3}.
\end{gathered}
\end{equation}
\end{subequations}
The states $\ket{n;\sigma_1\sigma_2;n^-_3,n^+_3}$
consist of two singlets and a doublet under $\vec{S}$
\footnote{The notation here should not be confused with that introduced at the 
beginning of Section~\ref{sec:QM:spectrum}, where fermion occupation
numbers were used to label states. Here, $\sigma_1$ and $\sigma_2$
label the supersymmetry multiplet.}; a multiplet of degeneracy $n$
under $\vec{N}_-$; and a multiplet of degeneracy $n-\abs{\kappa}$
under $\vec{N}_+$.  Thus, the total degeneracy of the energy
level~\eqref{2mE} is
\begin{equation} \label{deg}
\text{degeneracy} = 4 n (n-\abs{\kappa}).
\end{equation}
As in footnote~\ref{ft:deg} (p.~\pageref{ft:deg}),
it has really only been shown that the bound states
at level $n$
come in multiples of $4 n (n-\abs{\kappa})$, and in fact there are
no bound states if $\kappa \param > 0$, but there are indeed
precisely $4 n (n-\abs{\kappa})$ states with quantum number $n$ if
$\kappa \param<0$.

\subsubsection{Angular Momentum Basis of Nonsupersymmetric Bound States}
\label{sec:boundJspectrum}

The basis~\eqref{nbasis} is not very physical because angular momentum
is not well-defined in that basis.  This is easily fixed.  Because
\begin{equation} \label{JisSum}
\vec{J} = \Vec{\tilde{J}} + \vec{S} 
= \Vec{\tilde{N}}_+ + \Vec{\tilde{N}}_- + \vec{S},
\end{equation}
the basis adapted to the angular momentum is a simple application of
the rules for summing angular momentum.  In particular, since $\vec{S}$
does not commute with $\vec{J}$, it is not possible to specify both, but
it is still possible to simultaneously specify the fermion number with the
angular momentum.  The resulting states are therefore
\begin{equation} \label{JStateEnumeration}
\ket{n;\sigma_1+\sigma_2;\tj,j,j_3}, \qquad
\begin{aligned}
\tj &= \tfrac{\abs{\kappa}}{2}, \tfrac{\abs{\kappa}}{2} + 1, \dotsc,
            n-\tfrac{\abs{\kappa}}{2}-2, n-\tfrac{\abs{\kappa}}{2}-1, \\
j &= \begin{cases} \tj, &\sigma_1+\sigma_2 = 0,2, \\
                  \tj \pm \frac{1}{2}, & \sigma_1+\sigma_2 = 1,
\end{cases} \\
j_3 &= -j,-j+1,\dotsc,j-1,j,
\end{aligned}
\end{equation}
where $\sigma_1+\sigma_2$ is the fermion occupation number,
and $\Vec{\tilde{J}}^2 = \tj(\tj+1)$.

It is straightforward to use Clebsch-Gordan coefficients to find
the fermionic states from the bosonic ones.  Specifically,
\begin{subequations} \label{boundFAng}
\begin{align}
\begin{split}
\ket{n;1;j-\tfrac{1}{2},j,j_3} =&
\frac{1}{\sqrt{4 j E}} \left[
 \sqrt{j+j_3}\bar{Q}^1\ket{n;0;j-\tfrac{1}{2},j-\tfrac{1}{2},j_3-\tfrac{1}{2}} 
\right. \\ & \left.
 +\sqrt{j-j_3}\bar{Q}^2\ket{n;0;j-\tfrac{1}{2},j-\tfrac{1}{2},j_3+\tfrac{1}{2}}
\right],
\quad j = \tfrac{\abs{\kappa}+1}{2}, \tfrac{\abs{\kappa}+3}{2}, \dotsc,
            \skipthis{n-\tfrac{\abs{\kappa}+3}{2},}
            n-\tfrac{\abs{\kappa}+1}{2}.
\end{split} \\
\begin{split}
\ket{n;1;j+\tfrac{1}{2},j,j_3} =&
\frac{1}{\sqrt{4(j+1)E}}
\left[-
\sqrt{j-j_3+1}\bar{Q}^1\ket{n;0;j+\tfrac{1}{2},j+\tfrac{1}{2},j_3-\tfrac{1}{2}}
\right. \\ & \left. \mspace{-30mu}
+ \sqrt{j+j_3+1} \bar{Q}^2
  \ket{n;0;j+\tfrac{1}{2},j+\tfrac{1}{2},j_3+\tfrac{1}{2}}
\right],
\quad j = \tfrac{\abs{\kappa}-1}{2}, \tfrac{\abs{\kappa}+1}{2}, \dotsc,
            \skipthis{n-\tfrac{\abs{\kappa}+5}{2},}
             n-\tfrac{\abs{\kappa}+3}{2}.
\end{split}
\end{align}
\end{subequations}

\subsection{Marginally Bound States} \label{sec:marginspectrum}

When
$2mE = \param^2$, the algebra is the contraction
\begin{align} \label{alg:contract}
\com{\tilde{J}^i}{\tilde{J}^j} &= i \epsilon^{ijk} \tilde{J}^k, &
\com{\tilde{J}^i}{\tilde{K}^j} &= i \epsilon^{ijk} \tilde{K}^k, &
\com{\tilde{K}^i}{\tilde{K}^j} &= 0, && i=1,2,3.
\end{align}
Irreducible representations are characterized by the
Casimirs~\eqref{QMCasimirs}
\begin{align} \label{marginCasimirs}
\Vec{\tilde{K}}^2 &= \frac{\kappa^2\param^2}{4}, &
\Vec{\tilde{J}}\cdot\Vec{\tilde{K}} &= \frac{\kappa^2\param}{4},
\end{align}
and since $\Vec{\tilde{J}}$ and $\Vec{\tilde{K}}$ are hermitian,
it is unitary representations that are the representations of interest.

Because of the SU(2) subalgebra generated by $\Vec{\tilde{J}}$,
a general irreducible
unitary representation of the algebra~\eqref{alg:contract} consists
of a sum of SU(2) representations.
As reviewed in Appendix~\ref{sec:SO31}, because
$\Vec{\tilde{J}}\cdot \Vec{\tilde{K}} \neq 0$,
a unitary irreducible representation of the algebra~\eqref{alg:contract}
contains
exactly one copy of every SU(2)
representation $j=j_0, j_0+1, \dotsc$, where  $j_0 = 
\frac{\Vec{\tilde{J}}\cdot\Vec{\tilde{K}}}{\abs{\Vec{\tilde{K}}}}$.

Thus, for the case at hand, marginally bound states consist of the states
with $\Vec{\tilde{J}}^2 = \tj(\tj+1)$,
$\tj=\frac{\abs{\kappa}}{2},
\frac{\abs{\kappa}}{2}+1,
\dotsc$.  The same argument as in $\S$\ref{sec:boundJspectrum} then
gives the states
\begin{subequations} \label{marginalFAng}
\begin{gather} \label{marginalFAng:bosonic}
\begin{align}
\ket{E=\tfrac{\param^2}{2m};0;j,j,j_3}, &&
\ket{E=\tfrac{\param^2}{2m};2;j,j,j_3}
=& \frac{m}{\param^2} \bar{Q}^1 \bar{Q}^2
   \ket{E=\tfrac{\param^2}{2m};0;j,j,j_3}, &
j = \tfrac{\abs{\kappa}}{2}, \tfrac{\abs{\kappa}}{2}+1, \dotsc,
\end{align} \\ \label{marginalFAng:up}
\begin{align} 
\begin{split}
\ket{E=\tfrac{\param^2}{2m};1;j-\frac{1}{2},j,j_3}
=& \sqrt{\frac{m}{2 j \param^2}} \left[
 \sqrt{j+j_3} \bar{Q}^1 \ket{E=\tfrac{\param^2}{2m};0;j-\tfrac{1}{2},
                             j-\tfrac{1}{2},j_3-\tfrac{1}{2}} 
\right. \\ & \left. \mspace{-30mu}
 + \sqrt{j-j_3} \bar{Q}^2 
  \ket{E=\tfrac{\param^2}{2m};0;j-\tfrac{1}{2},j-\tfrac{1}{2},j_3+\tfrac{1}{2}}
\right],
\quad j = \tfrac{\abs{\kappa}+1}{2}, \tfrac{\abs{\kappa}+3}{2}, \dotsc,
\end{split}
\\
\begin{split}
\ket{E=\tfrac{\param^2}{2m};1;j+\tfrac{1}{2},j,j_3} =&
\sqrt{\frac{m}{2(j+1)\param^2}}
\left[-
\sqrt{j-j_3+1} \bar{Q}^1 
\ket{E=\tfrac{\param^2}{2m};0;j+\tfrac{1}{2},j+\tfrac{1}{2},j_3-\tfrac{1}{2}}
\right. \\ & \left. \mspace{-60mu}
+ \sqrt{j+j_3+1} \bar{Q}^2 
  \ket{E=\tfrac{\param^2}{2m};0;j+\tfrac{1}{2},j+\tfrac{1}{2},j_3+\tfrac{1}{2}}
\right], \quad j = \tfrac{\abs{\kappa}-1}{2},
                    \tfrac{\abs{\kappa}+1}{2}, \dotsc,
\end{split}
\end{align} \label{marginalFAng:down}
\end{gather}
\end{subequations}
for the marginally bound states in an angular momentum basis.

\subsection{Unbound States}

When
$2mE > \param^2$, the relevant algebra is SO(3,1),
\begin{equation} \label{alg:unbound}
\begin{gathered}
\begin{aligned}
\com{\tilde{J}^i}{\tilde{J}^j} &= i \epsilon^{ijk} \tilde{J}^k, &
\com{\tilde{J}^i}{\hat{K}^j} &= i \epsilon^{ijk} \hat{K}^k, &
\com{\hat{K}^i}{\hat{K}^j} &= -i \epsilon^{ijk} \tilde{J}^k, &
& i=1,2,3,
\end{aligned} \\
\Vec{\hat{K}} \equiv  \frac{1}{\sqrt{2 m E - \param^2}} \Vec{\tilde{K}}.
\end{gathered}
\end{equation}
Irreducible representations are characterized by the
Casimirs~\eqref{QMCasimirs}
\begin{align} \label{unboundCasimirs}
\Vec{\tilde{J}}^2 - 
\Vec{\hat{K}}^2 &= \frac{\kappa^2}{2} -1
  - \frac{\kappa^2}{2} \frac{m E}{2mE-\param^2}, &
\Vec{\tilde{J}}\cdot\Vec{\hat{K}} &= \frac{\kappa^2\param}{
   4\sqrt{2mE-\param^2}},
\end{align}
and since $\Vec{\tilde{J}}$ and $\Vec{\hat{K}}$ are hermitian
for $2mE > \param^2$,
it is unitary representations that are the representations of interest.

Because of the SU(2) subalgebra generated by $\Vec{\tilde{J}}$,
a general irreducible
unitary representation of the algebra~\eqref{alg:contract} consists
of a sum of SU(2) representations.
According to the
SO(3,1) representation theory reviewed in Appendix~\ref{sec:SO31},
the Casimirs of the relevant unitary irreducible representation
are
$\Vec{\tilde{J}}\cdot\Vec{\hat{K}} = \tj_0 \zeta$ and
$\Vec{\tilde{J}}^2-\Vec{\hat{K}}^2 = \zeta^2 + 1 - \tj_0^2$, where
$\tj_0$ is a half-integer which labels the minimal irreducible SU(2)
representation in the irreducible representation of the full
algebra~\eqref{alg:contract} and $\zeta$ is real.
The
irreducible representation contains exactly one copy of every SU(2)
representation with $\tj=\tj_0, \tj_0+1, \dotsc$.
In the case at hand, $\tj_0=\frac{\abs{\kappa}}{2}$, exactly as in
the marginally bound case~$\S$\ref{sec:marginspectrum},
with the same conclusions.

\section{Wavefunctions} \label{sec:states}

Now begins the task of constructing wavefunctions.
Roughly speaking, the wavefunctions can be separated into a radial part
and an angular part.  This is strictly true for the bosonic states---those
with even fermion number---and it is only slightly more complicated
for the fermionic states with fermion number one.  Moreover, the
fermionic states are found from the bosonic states using
Eqs.~\eqref{boundFAng} or~\eqref{marginalFAng}.

\subsection{Angular Momentum Eigenfunctions} \label{sec:JFStates}

Because the bosonic part of the Hamiltonian differs from that of
a charged particle in a magnetic monopole background only by radial terms,
the angular dependence of the bosonic states is given by the spherical
harmonics for a particle in a magnetic monopole background.
In polar coordinates
$\vec{x}=(r\sin\polar\cos\phi,r\sin\polar\sin\phi,r\cos\polar)$,
and restricting to the northern hemisphere ($\polar < \frac{\pi}{2}+\epsilon$),
these are~\cite{wuyang},
\begin{multline} \label{Yjm}
Y_{j,j_3}(\polar,\phi) = \frac{(-1)^{j-j_3}}{2^{j_3} \sqrt{4\pi}}
 \sqrt{\frac{(2j+1) (j+j_3)!(j-j_3)!}{(j+\frac{\kappa}{2})!(j-\frac{\kappa}{2})!}}
 P_{j-j_3}^{(j_3-\frac{\kappa}{2},j_3+\frac{\kappa}{2})}(\cos\polar)
 \sin^{j_3} \polar \cot^{\frac{\kappa}{2}} \tfrac{\polar}{2}
 e^{i (j_3-\frac{\kappa}{2}) \phi},
\\  \polar < \frac{\pi}{2}+\epsilon, \quad j_3 = -j,-j+1,\dotsc,j-1,j,
\end{multline}
where%
\footnote{It is understood here that
$\frac{1}{\Gamma(c)}{{_2}F_1}(-n,b;c;z)
\equiv \frac{n!}{\Gamma(b)}
  \sum_{p=0}^n (-1)^p \frac{\Gamma(b+p)}{p!(n-p)!\Gamma(c+p)} z^p$,
which
is well-defined even when $c$ is a nonpositive integer, provided
that $b$ is not also a nonpositive integer.  Since
\hbox{$(j_3-\frac{\kappa}{2})+(j_3+\frac{\kappa}{2})+(j-j_3)+1 = j+j_3+1 \geq 1$}, this
condition on $b$ is satisfied in this paper.
\label{ft:2F1Regularized}}
\begin{equation} \label{Pn}
P_n^{(\alpha,\beta)}(x) = \frac{\Gamma(n+\alpha+1)}{n! \Gamma(\alpha+1)}
  {_2}F_1(-n,\alpha+\beta+n+1;\alpha+1;\tfrac{1-x}{2})
\end{equation}
is the Jacobi polynomial, expressed as a hypergeometric function.
Notice that single-valuedness (in $\phi$) requires $j-\frac{\kappa}{2}\in \ZZ$;
normalizability then demands the bound
$j\geq \frac{\abs{\kappa}}{2}$, which was
observed in the bosonic theory both classically
[Eq.~\eqref{JBound}] and quantum mechanically [Eqs.~\eqref{JStateEnumeration}
and~\eqref{marginalFAng:bosonic}].

In the southern hemisphere ($\polar > \frac{\pi}{2}-\epsilon$) the
gauge transformation~\eqref{DA} yields
\begin{multline} \label{Yjm:southern}
Y_{j,j_3}(\polar,\phi) = \frac{(-1)^{j-j_3}}{2^{j_3} \sqrt{4\pi}}
 \sqrt{\frac{(2j+1) (j+j_3)!(j-j_3)!}{(j+\frac{\kappa}{2})!(j-\frac{\kappa}{2})!}}
 P_{j-j_3}^{(j_3-\frac{\kappa}{2},j_3+\frac{\kappa}{2})}(\cos\polar)
 \sin^{j_3} \polar \cot^{\frac{\kappa}{2}} \tfrac{\polar}{2}
 e^{i (j_3+\frac{\kappa}{2}) \phi},
\\  \polar > \frac{\pi}{2}-\epsilon, \quad j_3 = -j,-j+1,\dotsc,j-1,j.
\end{multline}
Because the only difference is in the sign in front of $\kappa$ in
the phase $e^{\pm i\frac{\kappa}{2} \phi}$,
we will, without further comment, take
$\epsilon$ only slightly less than $\frac{\pi}{2}$ so
that essentially only those points on the negative $z$ axis are not covered
by the patch in the upper ``hemi''sphere.  That is, we will be content
to use~\eqref{Yjm} for the angular dependence, with the understanding that
near the south pole,~\eqref{Yjm:southern} should be used instead.

The spherical harmonics~\eqref{Yjm} are easily verified by using
the identities~\eqref{Pn} and
$\frac{d}{dx}{_2}F_1(\alpha,\beta;\gamma;x)=\frac{\alpha\beta}{\gamma}
  {_2}F_1(\alpha,\beta;\gamma;x)$ to find
\begin{equation}\label{derP}
\frac{d}{dx} P_n^{(\alpha,\beta)}(x) = \frac{1}{2} (\alpha+\beta+n+1)
  P_{n-1}^{(\alpha+1,\beta+1)}(x).
\end{equation}
It is then straightforward to verify
\begin{subequations}
\begin{align}
J^+ Y_{j,j_3}(\polar,\phi) &= \sqrt{(j-j_3)(j+j_3+1)} Y_{j,j_3+1}(\polar,\phi), &
J^3 Y_{j,j_3}(\polar,\phi) &= j_3 Y_{j,j_3}(\polar,\phi).
\end{align}
With a little more work%
\footnote{
It may be useful to use Eq.~\eqref{Jid2}%
\iftoomuchdetail
\else.
\fi
\label{ft:Pid}}
one can also verify
\begin{equation}
J^- Y_{j,j_3}(\polar,\phi) = \sqrt{(j+j_3)(j-j_3+1)} Y_{j,j_3-1}(\polar,\phi).
\end{equation}
\end{subequations}
It is easily seen that $Y_{j,j}(\polar,\phi)$ is properly normalized; the
recursion then ensures that $Y_{j,j_3}(\polar,\phi)$ is.

The bosonic harmonics~\eqref{Yjm} yield fermionic harmonics upon adding spin
via the Clebsch-Gordan coefficients.  This results in
\begin{subequations} \label{Yjjm}
\begin{align} \label{Yjjm:up}
\fang{j-\tfrac{1}{2}}{j}{j_3} &=
\sqrt{\frac{j+j_3}{2j}} Y_{j-\frac{1}{2},j_3-\frac{1}{2}}(\polar,\phi)
  \ket{10}
+ \sqrt{\frac{j-j_3}{2j}} Y_{j-\frac{1}{2},j_3+\frac{1}{2}}(\polar,\phi)
  \ket{01}, \\ \label{Yjjm:down}
\fang{j+\tfrac{1}{2}}{j}{j_3} &=
-\sqrt{\frac{j-j_3+1}{2(j+1)}} Y_{j+\frac{1}{2},j_3-\frac{1}{2}}(\polar,\phi)
  \ket{10}
+ \sqrt{\frac{j+j_3+1}{2(j+1)}} Y_{j+\frac{1}{2},j_3+\frac{1}{2}}(\polar,\phi)
  \ket{01},
\end{align}
\end{subequations}
using the basis for fermionic states, Eq.~\eqref{Fstates}.
These states are single-valued if $j$ differs from $\frac{\kappa}{2}$ by
half an odd integer.  In particular, the
states~\eqref{Yjjm:up} are well-defined for
$0\leq j-\frac{\kappa+1}{2} \in \ZZ$
and the states~\eqref{Yjjm:down} are well-defined for
$0 \leq j-\frac{\kappa-1}{2} \in \ZZ$.

Please do not be fooled by the resemblance between the notation in the
definitions~\eqref{Yjjm} and~\eqref{boundFAng}.  The
states~\eqref{boundFAng} have a definite~$\Vec{\tilde{J}}^2$ and are
otherwise defined by the angular momentum
$\vec{J}=\Vec{\tilde{J}}+\Vec{S}$, but for the harmonics~\eqref{Yjjm},
the angular momentum has been decomposed as a sum of orbital and spin
angular momenta, ie. $\vec{J} = \vec{L} + \vec{s}$.  The former are
natural from a supersymmetric perspective, but the latter is more
physical.

\subsection{Ground State Wavefunctions} \label{sec:susyStates}

Because of the SU(2)$^2$ symmetry algebra, generated by
$\vec{N}_\pm$ [Eq.~\eqref{N+-}],
associated with the ground states, one ground state can be found by
demanding that it is annihilated by the raising operators $N^+_\pm$,
and that it be an eigenstate of $N^3_\pm$.
The remaining ground states are found by applying the lowering operators
$N^-_\pm$.

In fact, because of the Casimirs~\eqref{NCasimir},
the equations to solve are equivalently
\begin{subequations} \label{highestGround}
\begin{align} \label{highestGround:J}
J^+ \ket{E;\tfrac{\abs{\kappa}-1}{2},\tfrac{\abs{\kappa}-1}{2}} &= 0, &
J^3 \ket{E;\tfrac{\abs{\kappa}-1}{2},\tfrac{\abs{\kappa}-1}{2}} &=
  \tfrac{\abs{\kappa}-1}{2} \ket{E;\tfrac{\abs{\kappa}-1}{2},
                                 \tfrac{\abs{\kappa}-1}{2}}, \\
\label{highestGround:J3}
K^+ \ket{E;\tfrac{\abs{\kappa}-1}{2},\tfrac{\abs{\kappa}-1}{2}} &= 0, &
K^3 \ket{E;\tfrac{\abs{\kappa}-1}{2},\tfrac{\abs{\kappa}-1}{2}} &=
   \tfrac{\abs{\kappa}-1}{2} \ket{E;\tfrac{\abs{\kappa}-1}{2},
                                 \tfrac{\abs{\kappa}-1}{2}},
\end{align}
\end{subequations}
and moreover it was learned in $\S$\ref{sec:susyspectrum} that
these states have fermion number one.
But Eqs.~\eqref{highestGround:J} have already been solved, more generally,
by~Eq.\eqref{Yjjm:down}.  That is, in a coordinate basis, the ground
states are given by
\begin{equation}
\braket{r,\polar,\phi}{E;\tfrac{\abs{\kappa}-1}{2},j_3} =
-\sqrt{\tfrac{\abs{\kappa}+1-2j_3}{2(\abs{\kappa}+1)}} \radwav_g(r)
 Y_{\frac{\abs{\kappa}}{2},j_3-\frac{1}{2}}(\polar,\phi) \ket{10}
+ \sqrt{\tfrac{\abs{\kappa}+1+2j_3}{2(\abs{\kappa}+1)}} \radwav_g(r)
 Y_{\frac{\abs{\kappa}}{2},j_3+\frac{1}{2}}(\polar,\phi) \ket{01}.
\end{equation}
In particular, the state annihilated by $J^+$ and $K^+$ is
\begin{multline} \label{highestGround:allButRadial}
\braket{r,\polar,\phi}{E;\tfrac{\abs{\kappa}-1}{2},\tfrac{\abs{\kappa}-1}{2}} =
\sqrt{\tfrac{\abs{\kappa}}{4\pi}}
\radwav_g(r) \sin^{\frac{\abs{\kappa}-\kappa}{2}}\tfrac{\polar}{2}
          \cos^{\frac{\abs{\kappa}+\kappa}{2}}\tfrac{\polar}{2}
e^{i \frac{\abs{\kappa}-\kappa}{2} \phi}
\\ \times
\left[ \left(1-\tfrac{\kappa}{\abs{\kappa}}
 -2 \sin^2 \tfrac{\polar}{2} \right) \csc \polar e^{-i \phi}
 \ket{10}
+ \ket{01}
\right].
\end{multline}
It remains to determine the radial function
$\radwav_g(r)$.

The radial function is determined by
solving~$K^+\ket{E;\tfrac{\abs{\kappa}-1}{2},\tfrac{\abs{\kappa}-1}{2}}=0$.
A little algebra converts the component of this equation along
$\ket{01}$ into
\begin{equation}
\radwav_g'(r) = \frac{1}{r}\left(\frac{\abs{\kappa}}{2}-1\right) \radwav_g(r)
 + \frac{\kappa\param}{\abs{\kappa}} \radwav_g(r),
\end{equation}
and therefore
\begin{equation}
\radwav_g(r) = \frac{(2\abs{\param})^{3/2}}{\sqrt{(\abs{\kappa})!}}
  \left(2 \abs{\param} r\right)^{\frac{\abs{\kappa}}{2}-1}
  e^{\frac{\kappa}{\abs{\kappa}} \param r}.
\end{equation}
Notice that this is normalizable only if $\kappa \param < 0$, the same
condition found for the existence of classically bound orbits.

Applying the lowering operator $J^-$ thus yields the ground state
wavefunctions
\begin{multline}
\braket{r,\polar,\phi}{0;\tfrac{\abs{\kappa}-1}{2},j_3}
= (-1)^{\frac{\abs{\kappa}-1}{2}-j_3}
  \frac{2 \abs{\param}^{3/2}}{\sqrt{\pi} \abs{\kappa}!}
  \sqrt{\bigl(\tfrac{\abs{\kappa}-1}{2}+j_3\bigr)!
        \bigl(\tfrac{\abs{\kappa}-1}{2}-j_3\bigr)!}
\\ \times
\left(2 \abs{\param} r\right)^{\frac{\abs{\kappa}}{2}-1}
e^{\frac{\kappa}{\abs{\kappa}} \param r}
\sin^{j_3-\frac{1}{2}}\polar \, \cot^{\frac{\kappa}{2}}\tfrac{\polar}{2}
e^{i \frac{\abs{\kappa}-\kappa}{2} \phi}
\\ \times
\left[ P_{\frac{\abs{\kappa}+1}{2}-j_3}^{(j_3-\frac{\kappa+1}{2},
           j_3+\frac{\kappa-1}{2})}(\cos\polar) e^{-i \phi} \ket{10}
+ \frac{1}{2} \sin\polar 
   P_{\frac{\abs{\kappa}-1}{2}-j_3}^{(j_3-\frac{\kappa-1}{2},
           j_3+\frac{\kappa+1}{2})}(\cos\polar)\ket{01}
\right].
\end{multline}
Again, these states only exist if $\kappa\param<0$
and for $j_3 = -\tfrac{\abs{\kappa}-1}{2},\dotsc,\tfrac{\abs{\kappa}-1}{2}$.
There are $\abs{\kappa}$ of these states.

\subsection{Bound State Wavefunctions} \label{sec:boundStates}

In this section, we present the wavefunctions of the bound states.
Recall from $\S$\ref{sec:boundspectrum}
that the algebra of $\Vec{\Tilde{N}}_\pm$ gives the states
$\ket{n;\sigma_1,\sigma_2;n_3^-,n_3^+}$.  If these were the states of
interest, then operator techniques such as those in $\S$\ref{sec:susyStates}
would give the wavefunctions.  In particular,
$\ket{n;\sigma_1,\sigma_2;\frac{n-1}{2},\frac{n-\abs{\kappa}-1}{2}}$
is found by demanding
that it be annihilated by $\tilde{K}^+$ and $\tilde{J}^+$, and
that it have the correct $\tilde{K}_3$ and $\tilde{J}_3$ eigenvalues.

However, the relation of those states
to the states of~$\S$\ref{sec:boundJspectrum},
$\ket{n;\sigma_1+\sigma_2;\tj,j,j_3}$, is sufficiently
nontrivial, that a more brute force method will be applied.
We will use our knowledge of the angular part of the wavefunction
to find the equation of motion for the radial factor directly
from the Hamiltonian.  Fermionic wavefunctions are then found using
Eq.~\eqref{boundFAng}.

\subsubsection{Bosonic Bound States} \label{sec:boundBStates}

The wavefunctions of bosonic bound states will have the form
\begin{equation}
\bra{\sigma\sigma}\braket{r,\polar,\phi}{n;2\sigma;j,j,j_3}
 = \radwav_{n,j}(r) Y_{j,j_3}(\polar,\phi), \qquad
\sigma=0,1, \quad 
\begin{aligned}
j&=\tfrac{\abs{\kappa}}{2},\dotsc,n-\tfrac{\abs{\kappa}}{2}-1, \\
j_3&=-j,\dotsc,j.
\end{aligned}
\end{equation}
Given the energy eigenvalue~\eqref{2mE},
and writing the Hamiltonian~\eqref{H} in terms of $\vec{J}^2$ then yields
the equation
\begin{multline} \label{boundSchroedinger}
-\radwav_{n,j}''(r) - \frac{2}{r} \radwav_{n,j}'(r) + 
 \frac{j(j+1)}{r^2} \radwav_{n,j}(r)
+ \frac{\kappa\param}{r} \radwav_{n,j} +\param^2 \radwav_{n,j}(r)
= \frac{4 n \param^2 (n - \abs{\kappa})}{(2n-\abs{\kappa})^2}
   \radwav_{n,j}(r).
\end{multline}
Notice that the $\frac{\kappa^2}{8m r^2}$ part of the potential was absorbed
into $\frac{\vec{J}^2}{2mr^2}$ (and we have multiplied Schr\"{o}dinger's
equation by $2m$).
If $\kappa\param>0$, there are no regular solutions, as expected.%
\footnote{For $\kappa\param>0$, the putative solution would
be 
\begin{equation}
\radwav_{n,j}(r) \propto r^j L_{-n-j+\frac{\abs{\kappa}}{2}-1}^{2j+1}\bigl(
   \tfrac{2\param\kappa}{2n-\abs{\kappa}}\bigr)
      e^{-\frac{\param\kappa}{2n-\abs{\kappa}} r}.
\end{equation}
However, the index of the Laguerre polynomial is negative.  Thus, either
this expression is zero, or the proportionality constant is infinite, 
whereupon the ``polynomial'' is an infinite series whose asymptotic
behaviour is $e^{\frac{2\param \kappa}{2n-\abs{\kappa}} r}$
and so leads to a divergent function.
\label{ft:Laguerre}}
Otherwise,
for $\kappa\param<0$,
the regular solution is given in terms of an
associated Laguerre polynomial,
$L_n^k(x)$,
\begin{equation} \label{radialnj}
\radwav_{n,j}(r) = 
 \frac{(-2\param\kappa)^{3/2}}{(2n-\abs{\kappa})^2}
\sqrt{\frac{(n-j-\frac{\abs{\kappa}}{2}-1)!}{
  (n+j-\frac{\abs{\kappa}}{2})!}}
\left(-\frac{2\param\kappa}{2n-\abs{\kappa}}r\right)^j 
L_{n-j-\frac{\abs{\kappa}}{2}-1}^{2j+1}\bigl(
   - \tfrac{2\param\kappa}{2n-\abs{\kappa}}r\bigr)
      e^{\frac{\param\kappa}{2n-\abs{\kappa}} r}.
\end{equation}

Thus, the bound (but excited) state bosonic wavefunctions are
\begin{multline} \label{bosonicbound}
\hspace{-6pt}
\bra{\sigma\sigma}\braket{r,\polar,\phi}{n;2\sigma;j,j,j_3}
=
\frac{(-1)^{j-j_3}(-2\param\kappa)^{3/2}}{
 2^{j_3}(2n-\abs{\kappa})^2\sqrt{4\pi}}
\sqrt{\frac{(2j+1) (n-j-\frac{\abs{\kappa}}{2}-1)!(j+j_3)!(j-j_3)!}{
  (n+j-\frac{\abs{\kappa}}{2})!
  (j+\frac{\kappa}{2})!(j-\frac{\kappa}{2})!}}
\\ \times
\left(-\frac{2\param\kappa}{2n-\abs{\kappa}}r\right)^j 
L_{n-j-\frac{\abs{\kappa}}{2}-1}^{2j+1}\bigl(
   - \tfrac{2\param\kappa}{2n-\abs{\kappa}}r\bigr)
      e^{\frac{\param\kappa}{2n-\abs{\kappa}} r}
 P_{j-j_3}^{(j_3-\frac{\kappa}{2},j_3+\frac{\kappa}{2})}(\cos\polar)
 \sin^{j_3} \polar \cot^{\frac{\kappa}{2}} \tfrac{\polar}{2}
 e^{i (j_3-\frac{\kappa}{2}) \phi}.
\end{multline}
This result might have been partially anticipated if one reasoned that
since the classical trajectories are conic sections, one should get
radial dependence similar to the hydrogen atom; however, it is not
obvious that this should be true since the plane of classical
trajectories is off of the origin.

\subsubsection{Fermionic Bound States} \label{sec:boundFStates}

Fermionic states are found from the bosonic ones,
Eq.~\eqref{bosonicbound} with $\sigma=0$, using
Eq.~\eqref{boundFAng}.\footnote{See Appendix~\ref{ap:fermWave}.}
By matching powers of $x$
it is possible to demonstrate that
\begin{equation}
{L'}^{(\alpha)}_n(x) = -\frac{n}{\alpha+1}L_n^{(\alpha)}(x) 
                - \frac{x}{\alpha+1}L_{n-1}^{(\alpha+2)}(x),
\end{equation}
and therefore
\begin{equation}\begin{split}
{\radwav'}_{n, j}(r) =& \left(\frac{j}{r} + \frac{\param\kappa}{2(j+1)}\right)\radwav_{n, j}(r)
                        + \frac{\param\kappa}{2n-|\kappa|}
                        \sqrt{\frac{(2n-|\kappa|)^2}{4(j+1)^2}-1}\,\radwav_{n, j+1}(r).
\end{split}\end{equation}
Thus, using Eq.~\eqref{fermWavefcns} one can write the bound state
fermionic wavefunctions in the form
\begin{tinyeq}
\begin{subequations}
\begin{align}
\begin{split}
&\braket{r,\polar,\phi}{n;1;j-\tfrac{1}{2},j,j_3} = \frac{1}{\sqrt{n(n-|\kappa|)}}\Bigg\{
        \frac{\frac{\kappa(2n-|\kappa|)}{2j+1} \radwav_{n,j-\frac{1}{2}}(r) + \kappa
                \sqrt{\left(\frac{2n-|\kappa|}{2j+1}\right)^2 - 1} \radwav_{n,j+\frac{1}{2}}(r)}{2j+1}
                \sqrt{(j+\tfrac{1}{2})^2 - \tfrac{\kappa^2}{4}}\fang{j+\tfrac{1}{2}}{j}{j_3}\\
        &\quad +\left(
                \kappa\frac{\frac{\kappa(2n-|\kappa|)}{2j+1} \radwav_{n,j-\frac{1}{2}}(r) + \kappa
                \sqrt{\left(\frac{2n-|\kappa|}{2j+1}\right)^2 - 1} \radwav_{n,j+\frac{1}{2}}(r)}{2(2j+1)}
                -\frac{2n-|\kappa|}{2}\radwav_{n,j-\frac{1}{2}}(r)
                \right)\fang{j-\tfrac{1}{2}}{j}{j_3}\Bigg\},
\end{split}\raisetag{4\baselineskip}\normalTag\\
\begin{split}
&\braket{r,\polar,\phi}{n;1;j+\tfrac{1}{2},j,j_3} = \frac{1}{\sqrt{n(n-|\kappa|)}}\Bigg\{\\
        &\quad \frac{\left(\frac{2(j+1)}{\param r} + \frac{\kappa}{2j+3}\right)(2n-|\kappa|)\radwav_{n,j+\frac{1}{2}}(r) 
                + \kappa\sqrt{\left(\frac{2n-|\kappa|}{2j+3}\right)^2-1}\radwav_{n, j+\frac{3}{2}}(r)}{2j+1}
                        \sqrt{(j+\tfrac{1}{2})^2-\tfrac{\kappa^2}{4}}\fang{j-\tfrac{1}{2}}{j}{j_3}\\
        &\quad - \left(
         \kappa\frac{\left(\frac{2(j+1)}{\param r} + \frac{\kappa}{2j+3}\right)(2n-|\kappa|)\radwav_{n,j+\frac{1}{2}}(r) 
                + \kappa\sqrt{\left(\frac{2n-|\kappa|}{2j+3}\right)^2-1}\radwav_{n, j+\frac{3}{2}}(r)}{2(2j+1)}
                +\frac{2n-|\kappa|}{2} \radwav_{n,j+\frac{1}{2}}(r)\right)\fang{j+\tfrac{1}{2}}{j}{j_3}\Bigg\}.
\end{split}
\normalTag
\end{align}
\end{subequations}
\end{tinyeq}

\subsection{Unbound States} \label{sec:unboundStates}

\subsubsection{Bosonic Unbound States}
Replacing the eigenvalue on the right hand side of
the radial equation~\eqref{boundSchroedinger} with $2mE - \param^2$
immediately leads to the radial eigenfunctions
\begin{equation} \label{bosonicNBWF}
\radwav_{E,j}(r) =
\frac{\sqrt{2mE - \param^2}
\abs{\Gamma(j+1+i \frac{\kappa\param}{2 \sqrt{2mE-\param^2}})}}{
  2\sqrt{2\pi}(2j+1)! m}
e^{\frac{-\pi\kappa\param}{4\sqrt{2mE-\param^2}}}
\frac{M_{-i \frac{\kappa \param}{2\sqrt{2mE-\param^2}},j+\frac{1}{2}}
 \bigl(-2i \sqrt{2mE-\param^2} r\bigr)}{r},
\end{equation}
where $M_{k,\mu}(x)$ is the Whittaker function,
\begin{equation}
M_{k,\mu}(x) = e^{-x/2} x^{\mu+\frac{1}{2}} {_1F_1}(\mu-k+\tfrac{1}{2};
  2\mu+1;x).
\end{equation}
The confluent hypergeometric function which appears might also be called a 
nonpolynomial Laguerre function.

In the wavefunction~\eqref{bosonicNBWF} a particular square root of
$(2mE-\param^2)$ has been chosen; the
identity~\cite[Eq.~(9.231.2)]{gr} shows that the wavefunction with the
other square root differs from this one by at most a sign.
The normalization has
been chosen so that
\begin{equation}
\int_0^\infty dr r^2 \radwav^*_{E',j}(r) \radwav_{E,j}(r)
= \delta(E-E'),
\end{equation}
thus ensuring orthonormal wavefunctions.
See appendix~\ref{sec:norm}.

\subsubsection{Fermionic Unbound States}

Finding the fermionic unbound state wavefunctions is a simple matter of
applying Eq.~\eqref{fermWavefcns} for the radial dependence in
Eq.~\eqref{bosonicNBWF}. Using the identity
\begin{equation}
M'_{k, \mu}(x)= \frac{1}{x}(\mu+\tfrac{1}{2})M_{k, \mu}(x) 
        -\frac{k}{2\mu+1}M_{k, \mu}(x) 
  + \frac{(\mu+\tfrac{1}{2})^2-k^2}{2(2\mu + 1)^2(\mu+1)}M_{k, \mu+1}(x),
\end{equation}
which can be derived from the confluent hypergeometric representation, it
is possible to write
\begin{equation}
\radwav'_{E, j}(r) = \left(\frac{j}{r}-\frac{\kappa\param}{2(j+1)}\right)\radwav_{E, j}(r)
                - i\sqrt{2mE-\param^2 
                + \frac{\kappa^2\param^2}{4(j+1)^2}}\radwav_{E, j+1}(r).
\end{equation}
Plugging into Eq.~\eqref{fermWavefcns}, one finds the fermionic
wavefunctions in terms of the bosonic radial dependence and the
fermionic harmonics
\begin{tinyeq}
\begin{subequations}
\begin{align}
\begin{split}
&\braket{r, \polar, \phi}{E; 1; j-\tfrac{1}{2}, j, j_3}
= -\frac{1}{\sqrt{2mE}}\Biggl\{
\frac{i\sqrt{2mE-\param^2+\frac{\kappa^2\param^2}{(2j+1)^2}}
        \radwav_{E, j+\frac{1}{2}}(r)+
        \frac{\kappa\param}{2j+1}\radwav_{E, j-\frac{1}{2}}(r)}
        {j+1}\sqrt{(j+\tfrac{1}{2})^2-\tfrac{\kappa^2}{4}}
        \fang{j+\tfrac{1}{2}}{j}{j_3}\\
        &\qquad +\left(
        \kappa\frac{i\sqrt{2mE-\param^2+\frac{\kappa^2\param^2}{(2j+1)^2}}
        \radwav_{E, j+\frac{1}{2}}(r)+
        \frac{\kappa\param}{2j+1}\radwav_{E, j-\frac{1}{2}}(r)}
        {2(j+1)} +\param \radwav_{E, j-\frac{1}{2}}(r)
\right)\fang{j-\tfrac{1}{2}}{j}{j_3}
\Biggr\},
\end{split}\raisetag{2\baselineskip}\normalTag\\
\begin{split}\raisetag{2\baselineskip}
&\braket{r, \polar, \phi}{E; 1; j+\tfrac{1}{2}, j, j_3} = \frac{1}{\sqrt{2mE}}\Bigg\{
4\frac{\left(\frac{j+1}{r}-\frac{\kappa\param}{2j+3}\right)\radwav_{E, j+\frac{1}{2}}(r)
        -i\sqrt{2mE-\param^2+\frac{\kappa^2\param^2}{(2j+3)^2}}\radwav_{E, j+\frac{3}{2}}(r)}
                {2j+1}
        \sqrt{(j+\tfrac{1}{2})^2-\tfrac{\kappa^2}{4}}
        \fang{j-\tfrac{1}{2}}{j}{j_3}\\
&\quad - \left(2\kappa
        \frac{\left(\frac{j+1}{r}-\frac{\kappa\param}{2j+3}\right)\radwav_{E, j+\frac{1}{2}}(r)
        -i\sqrt{2mE-\param^2+\frac{\kappa^2\param^2}{(2j+3)^2}}\radwav_{E, j+\frac{3}{2}}(r)}
                {2j+1}
                +\param \radwav_{E, j+\frac{1}{2}}(r)\right)\fang{j+\tfrac{1}{2}}{j}{j_3}
\Bigg\}.
\end{split}
\normalTag
\end{align}
\end{subequations}
\end{tinyeq}

\subsection{Marginally Bound States} \label{sec:marginstate}

\subsubsection{Bosonic States}

The radial dependence for the marginally bound states with
$2mE=\param^2$ is
\begin{equation}
\radwav_{E=\frac{\param^2}{2m},j}(r)
\propto \frac{1}{\sqrt{-\kappa\param r}}J_{2j+1}\bigl(
2\sqrt{-\kappa\param r}\bigr),
\end{equation}
which is well-behaved at $r=\infty$ if $\kappa$ and $\param$ have opposite
sign.  The proper normalization is a $\delta$-function
normalization with the unbound states.
To be precise, we note that evaluating the radial wavefunction of
the unbound states near $2mE=\param^2$ gives
\begin{equation}
\radwav_{E\approx\frac{\param^2}{2m},j}(r)
\approx \frac{(2mE-\param^2)^{3/4}}{m} \sqrt{\frac{-\kappa\param}{2}}
e^{-\frac{\pi(\kappa\param+\abs{\kappa\param})}{4\sqrt{2mE-\param^2}}}
\frac{J_{2j+1}\bigl(2\sqrt{-\kappa\param r}\bigr)}{\sqrt{-\kappa\param r}},
\end{equation}
after dropping an irrelevant phase.  Notice that the exponential
factor is trivial when $\kappa\param<0$, but forces the threshold
wavefunction to strongly vanish for $\kappa\param>0$, in accord with
the presence or absence of bound states.

\subsubsection{Fermionic States}

Applying a standard Bessel function
identity~\cite[Eq.~(9.1.27)]{abramowitz}, one can demonstrate that
\begin{equation}
{\radwav'}_{E=\frac{\param^2}{2m}, j}(r) = -\frac{\radwav_{E=\frac{\param^2}{2m}, j}(r)}{2r}
        +\frac{1}{2}\sqrt{-\frac{\kappa\param}{r}}\radwav_{E=\frac{\param^2}{2m}, j-\frac{1}{2}}(r)
        -\frac{1}{2}\sqrt{-\frac{\kappa\param}{r}}\radwav_{E=\frac{\param^2}{2m}, j+\frac{1}{2}}(r),
\end{equation}
from which one finds the fermionic wave functions in terms of the
bosonic radial dependence and the fermionic harmonics using
Eq.~\eqref{fermWavefcns}:
\begin{tinyeq}
\begin{subequations}
\begin{align}
\begin{split}
&\braket{r, \polar, \phi}{E=\tfrac{\param^2}{2m}; 1; j-\tfrac{1}{2}, j, j_3} = 
                                                                \frac{1}{\param}\Bigg\{
 \frac{
                        \sqrt{\frac{-\kappa\param}{r}}
                        \left( \radwav_{ j-1}(r) - \radwav_{j}(r)\right) 
                        -\frac{2j}{r}\radwav_{ j-\frac{1}{2}}(r)}{2j+1}
                \sqrt{(j+\tfrac{1}{2})^2-\tfrac{\kappa^2}{4}}\fang{j+\tfrac{1}{2}}{j}{j_3}\\
        &\qquad+\left(\kappa\frac{
                        \frac{1}{2}\sqrt{\frac{-\kappa\param}{r}}
                        \left(\radwav_{ j-1}(r) - \radwav_{ j}(r)\right) 
                        -\frac{j}{r}\radwav_{j-\frac{1}{2}}(r)}{2j+1}
                -\param \radwav_{j-\frac{1}{2}}(r)
        \right)\fang{j-\tfrac{1}{2}}{j}{j_3}\Bigg\},
\end{split}\normalTag\\
\begin{split}
&\braket{r, \polar, \phi}{E=\tfrac{\param^2}{2m}; 1; j+\tfrac{1}{2}. j, j_3} = 
                                                                \frac{1}{\param}\Bigg\{ 
         \frac{\sqrt{\frac{-\kappa\param}{r}}\left(\radwav_{ j}(r)
                                - \radwav_{ j+1}(r)\right) 
                        + \frac{2(j+1)}{r}\radwav_{ j+\frac{1}{2}}(r)}{2j+1}
                \sqrt{(j+\tfrac{1}{2})^2-\tfrac{\kappa^2}{4}}\fang{j-\tfrac{1}{2}}{j}{j_3} \\
        &\qquad - \left(
                \kappa\frac{
                        \frac{1}{2}\sqrt{\frac{-\kappa\param}{r}}\left(\radwav_{j}(r)
                                - \radwav_{j+1}(r)\right)
                        +\frac{j+1}{r}\radwav_{j+\frac{1}{2}}(r)}{2j+1}
                +\param \radwav_{j+\frac{1}{2}}(r)
                \right)\fang{j+\tfrac{1}{2}}{j}{j_3}\Bigg\},
\end{split}
\normalTag
\end{align}
\end{subequations}
\end{tinyeq}
where, as should be clear, we have dropped the $E=\frac{\param^2}{2m}$
subscript on all of the above $\radwav$'s.

\acknowledgments

We thank B.~Chowdhury, S.~Mathur and S.~Nussinov for useful conversations.
We also thank M.~Headrick for making his useful and instructive
{\em Mathematica\/}$^\circledR$
package 
\href{http://www.stanford.edu/~headrick/physics/grassmann.m}{\tt grassmann.m\/}
available.
This work was supported in part by Department of Energy
contract No.~DE-FG02-91ER-40690.

\appendix

\section{Conventions} \label{sec:conventions}

Indexed bosonic coordinates $\vec{x} = (x^1, x^2, x^3)$ and
coordinates $\vec{x}=(x,y,z)$ are used interchangeably.  Overdots
denote time derivatives: $\Dot{\vec{x}}\equiv \frac{d \vec{x}}{dt}$.

Spinor fields
$\lambda,\xi$ are two-component spinors
whose conjugates are $\bar{\lambda}$ and $\bar{\xi}$.
If needed, the spinor indices $\alpha,\beta,\dotsi=1,2$
will be such that $\bar{\lambda}^\alpha = (\lambda_\alpha)^*$, so
that $\bar{\lambda} \lambda \equiv \bar{\lambda}^\alpha \lambda_\alpha$
is a scalar.  In this paper, we never need to contract
two unbarred or two barred spinors, and so we do not specify a convention
for this unnecessary operation.

The standard Pauli matrices are $\sigma^i$, or $\sigma^i_\alpha{^\beta}$.
In addition we use the standard raising and lowering operators
\begin{equation}
\sigma^\pm = \frac{1}{2} (\sigma^1 \pm i \sigma^2),
\end{equation}
so that $(\sigma^+) =
\begin{spmatrix} 0 & 1 \\ 0 & 0\end{spmatrix} = (\sigma^-)^\transpose$.
{\em For all other vectors and one-forms, the $\pm$ components do not
have a factor of $\frac{1}{2}$.}  For example,
\begin{align}
x^+ &= x^1 + i x^2 = x + i y, &
p_+ &= p_1 + i p_2, &
A_- &= A_1 - i A_2.
\end{align}

\section{Fermionic Wavefunctions}\label{ap:fermWave}

In this section, the general relationship of fermionic wavefunctions
to their bosonic superpartner is derived. The linear combinations of
the supersymmetry charges given in Eq.~\eqref{boundFAng} show how to
find the fermionic states from the bosonic states; however, it is not
a trivial matter to write the resulting fermionic wavefunctions in
terms of the more physically interpretable fermionic harmonics in
Eq.~\eqref{Yjjm}. In fact, it requires a number of Jacobi polynomial
identities or equivalently bosonic harmonic identities.

In general the bosonic wavefunctions may be written in the form
\begin{equation}
\braket{r, \polar,\phi}{E; 0; j, j, j_3} = \radwav_{E, j}(r) Y_{j ,j_3}(\polar, \phi),
\end{equation}
where the angular dependence is given in Eq.~\eqref{Yjm}. We
begin by applying the position space representations of the
supersymmetry charges in the combinations given in
Eq.~\eqref{boundFAng} to find
\begin{subequations}\label{fermPosSpace}
\begin{align}
&\braket{r, \polar, \phi}{E; 1; j-\tfrac{1}{2}, j, j_3}\notag\\
\begin{split} 
        &\qquad = \frac{1}{\sqrt{4jmE}}\Bigg\{
        \sqrt{j+j_3}
        \Big[ \left(\cos\polar\frac{\radwav'_-}{\radwav_-}- \frac{\sin\polar}{r} \frac{Y_{--}'}{Y_{--}}
                -\left(\frac{\kappa}{2r} + \param\right)\right)\ket{10}\\
        &\qquad\qquad\qquad\quad + \left(\sin\polar \frac{\radwav'_-}{\radwav_-} 
                        + \frac{\cos\polar}{r}\frac{Y_{--}'}{Y_{--}}
                + \frac{\frac{\kappa}{2}\cos\polar - j_3+\frac{1}{2}}{r\sin\polar}\right)e^{i\phi}\ket{01}
                \Big]\radwav_{-}Y_{--}\\ 
        &\qquad\qquad + \sqrt{j-j_3} \Big[
                -\left(\cos\polar \frac{\radwav'_-}{\radwav_-} - \frac{\sin\polar}{r}\frac{Y_{-+}'}{Y_{-+}}
                        + \left(\frac{\kappa}{2r} + \param\right)\right)\ket{01}\\
        &\qquad\qquad\qquad\quad + \left(\sin\polar \frac{\radwav'_-}{\radwav_-} 
                        + \frac{\cos\polar}{r}\frac{Y_{-+}'}{Y_{-+}}
                - \frac{\frac{\kappa}{2}\cos\polar - j_3-\frac{1}{2}}{r\sin\polar}\right)
                        e^{-i\phi}\ket{10}
                \Big]\radwav_-Y_{-+}
        \Bigg\},
\end{split}\\
&\braket{r,\polar, \phi}{E; 1;j+\tfrac{1}{2}, j, j_3}\notag\\
\begin{split}
        &\qquad = \frac{1}{\sqrt{4(j+1)mE}}\Bigg\{
        -\sqrt{j-j_3+1}
        \Big[ \left(\cos\polar\frac{\radwav'_+}{\radwav_+}- \frac{\sin\polar}{r} \frac{Y_{+-}'}{Y_{+-}}
                -\left(\frac{\kappa}{2r} + \param\right)\right)\ket{10}\\
        &\qquad\qquad\qquad\quad + \left(\sin\polar \frac{\radwav'_+}{\radwav_+} 
                        + \frac{\cos\polar}{r}\frac{Y_{+-}'}{Y_{+-}}
                + \frac{\frac{\kappa}{2}\cos\polar - j_3+\frac{1}{2}}{r\sin\polar}\right)e^{i\phi}\ket{01}
                \Big]\radwav_+Y_{+-}\\ 
        &\qquad\qquad + \sqrt{j+j_3+1} \Big[
                -\left(\cos\polar \frac{\radwav'_+}{\radwav_+} - \frac{\sin\polar}{r}\frac{Y_{++}'}{Y_{++}}
                        + \left(\frac{\kappa}{2r} + \param\right)\right)\ket{01}\\
        &\qquad\qquad\qquad\quad + \left(\sin\polar \frac{\radwav'_+}{\radwav_+} 
                        + \frac{\cos\polar}{r}\frac{Y_{++}'}{Y_{++}}
                - \frac{\frac{\kappa}{2}\cos\polar - j_3-\frac{1}{2}}{r\sin\polar}\right)
                        e^{-i\phi}\ket{10}
                \Big]\radwav_+Y_{++}
        \Bigg\},
\end{split}
\end{align}
\end{subequations}
with a shorthand
\begin{align}
\radwav_\pm &= \radwav_{E, j\pm \frac{1}{2}}, &
Y_{\pm\pm'} &= Y_{j\pm\frac{1}{2}, j_3\pm'\frac{1}{2}},
\end{align}
and where prime denotes differentiation with respect to $r$ or $\polar$,
depending on the context. To simplify these expressions we need to
relate $\p_\polar Y_{j, j_3}$ to other monopole
harmonics. Fortunately, this is not difficult with the identity
Eq.~\eqref{derP}. Once the expression is entirely in terms of the
bosonic harmonics there is a sizeable amount of algebraic manipulation
required to put the fermionic wavefunctions in terms of the fermionic
harmonics in Eq.~\eqref{Yjjm}. The relevant Jacobi polynomial
identities needed to effect this transformation are
\begin{subequations} \label{Jids}
\begin{align} \label{Jid1}
\frac{d}{dx} P_n^{(\alpha,\beta)}(x) =& \frac{1}{2} (\alpha+\beta+n+1)
        P_{n-1}^{(\alpha+1,\beta+1)}(x),\\
\begin{split}
-(n+1)P_{n+1}^{(\alpha-1, \beta-1)}(x) =&
        \frac{1}{4}(n+\alpha+\beta+1)(1-x^2)P_{n-1}^{(\alpha+1,\beta+1)}(x)\\
        &+\frac{1}{2}(\beta-\alpha-(\alpha+\beta)x)
        P_n^{(\alpha,\beta)}(x),\label{Jid2}
\end{split}\\ \label{Jid3}
\begin{split}
x P_n^{(\alpha, \beta)}(x) =& 
        \frac{2(n+1)(n+\alpha + \beta + 1)}{(2n+\alpha+\beta+1)(2n+\alpha+\beta+2)}
                 P_{n+1}^{(\alpha, \beta)}(x)\\
        &+ \frac{\beta^2-\alpha^2}{(2n+\alpha+\beta)(2n+\alpha+\beta+2)}P_n^{(\alpha, \beta)}(x)\\
        &+ \frac{2(n+\alpha)(n+\beta)}{(2n+\alpha+\beta)(2n+\alpha+\beta+1)}
                P_{n-1}^{(\alpha, \beta)}(x),
\end{split}\\ \label{Jid4}
\begin{split}
(1-x^2)P_n^{(\alpha, \beta)}(x) =& \frac{4(n+\alpha)(n+\beta)}{(2n+\alpha+\beta)(2n+\alpha+\beta+1)}P_n^{(\alpha-1, \beta-1)}(x)\\
                        &-\frac{4(\beta-\alpha)(n+1)}{(2n+\alpha+\beta)(2n+\alpha+\beta+2)}P_{n+1}^{(\alpha-1, \beta-1)}(x)\\
                        &-\frac{4(n+1)(n+2)}{(2n+\alpha+\beta+1)(2n+\alpha+\beta+2)}P_{n+2}^{(\alpha-1,\beta-1)}(x),
\end{split}\\ \label{Jid5}
\begin{split}
P_n^{(\alpha, \beta)}(x) =& 
        \frac{(n+\alpha+\beta+2)(n+\alpha+\beta+1)}{(2n+\alpha+\beta+2)(2n+\alpha+\beta+1)}
                P_n^{(\alpha+1, \beta+1)}(x)\\
        & -\frac{(\beta-\alpha)(n+\alpha+\beta+1)}{(2n+\alpha+\beta)(2n+\alpha+\beta+2)}
                P_{n-1}^{(\alpha+1, \beta+1)}(x)\\
        & -\frac{(n+\alpha)(n+\beta)}{(2n+\alpha+\beta)(2n+\alpha+\beta+1)}
                P_{n-2}^{(\alpha+1, \beta+1)}(x).
\end{split}
\end{align}
\end{subequations}
The identity~\eqref{Jid1} is explained just above Eq.~\eqref{derP} in
the main text. The identities~\eqref{Jid2} and~\eqref{Jid4} can be
derived by matching powers of $\tfrac{1-x}{2}$ in the hypergeometric
representation of the Jacobi polynomial, Eq.~\eqref{Pn}. The third
identity, Eq.~\eqref{Jid3}, is~\cite[Eq.~(22.7.1)]{abramowitz}. The
last identity, Eq.~\eqref{Jid5} can be derived using
\begin{equation}
\begin{split}
P_n^{(\alpha, \beta)}(x) &=
\frac{\Gamma(n+\alpha+1)}{n!\Gamma(\alpha+1)}
   {_2F_1}\bigl(-n,\alpha+\beta+n+1;\alpha+1;\tfrac{1-x}{2}\bigr)\\
&= \frac{\Gamma(n+\alpha+1)}{\Gamma(n+\alpha+\beta+1)}
\sum_{p=0}^n (-1)^p \frac{\Gamma(\alpha+\beta+n+1+p)}{p!(n-p)!
  \Gamma(\alpha+1+p)} \bigl(\tfrac{1-x}{2}\bigr)^p \\
&= \frac{\Gamma(n+\alpha+1)}{2^n \Gamma(n+\alpha+\beta+1)} (x-1)^n
\sum_{p=0}^n \frac{2^p \Gamma(\alpha+\beta+2n+1-p)}{p!(n-p)!
  \Gamma(\alpha+1+n-p)(x-1)^p},
\end{split}
\end{equation}
and matching powers of $(x-1)$.
The Jacobi polynomial identities~\eqref{Jids} translate into
the harmonic identities
\begin{subequations}
\begin{align}
\p_\polar Y_{j, j_3}(\polar, \phi)
        =& \frac{\tfrac{\kappa}{2}-j_3\cos\polar}{\sin\polar}Y_{j, j_3}(\polar, \phi)
         -\sqrt{(j+j_3)(j-j_3+1)}e^{i\phi}Y_{j, j_3-1}(\polar,\phi),\\
\p_\polar Y_{j, j_3}(\polar, \phi)
        =& \frac{j_3\cos\polar-\tfrac{\kappa}{2}}{\sin\polar}Y_{j, j_3}(\polar, \phi)
          +\!\!\sqrt{(j+j_3+1)(j-j_3)}e^{-i\phi}Y_{j, j_3+1}(\polar, \phi),\\
\begin{split}
\cos\polar\, Y_{j, j_3}(\polar, \phi) =& -\frac{1}{2(j+1)}\sqrt{\frac{
        ((j+1)^2-\frac{\kappa^2}{4})((j+1)^2-j_3^2)}
        {(j+\frac{1}{2})(j+\frac{3}{2})}}
        \, Y_{j+1, j_3}(\polar, \phi)\\
        & + \frac{\kappa j_3}{2j(j+1)}Y_{j, j_3}(\polar, \phi)\\
        &-\frac{1}{2j}\sqrt{\frac{(j+\frac{\kappa}{2})(j-\frac{\kappa}{2})(j+j_3)(j-j_3)}
                {(j-\frac{1}{2})(j+\frac{1}{2})}}\, Y_{j-1, j_3}(\polar, \phi),
\end{split}\\
\begin{split}
\sin\polar\, e^{-i\phi}Y_{j, j_3}(\polar, \phi) =& -\frac{1}{2(j+1)}\sqrt{\frac{
        ((j+1)^2-\frac{\kappa^2}{4})(j-j_3+2)(j-j_3+1)}{(j+\frac{1}{2})(j+\frac{3}{2})}}
                \,Y_{j+1, j_3-1}(\polar, \phi)\\
        & + \frac{\kappa}{2j(j+1)}\sqrt{(j+j_3)(j-j_3+1)}Y_{j, j_3-1}(\polar, \phi)\\
        & + \frac{1}{2j}\sqrt{\frac{(j+\frac{\kappa}{2})(j-\frac{\kappa}{2})(j+j_3)(j+j_3-1)}
                {(j-\frac{1}{2})(j+\frac{1}{2})}}\,Y_{j-1, j_3-1}(\polar, \phi),
\end{split}\raisetag{2\baselineskip}\\
\begin{split}
\sin\polar\, e^{i\phi}Y_{j, j_3}(\polar, \phi) =& \frac{1}{2(j+1)}\sqrt{\frac{
                ((j+1)^2 -\frac{\kappa^2}{4})(j+j_3+2)(j+j_3+1)}
                {(j+\frac{1}{2})(j+\frac{3}{2})}}\, Y_{j+1, j_3+1}(\polar, \phi)\\
        & +\frac{\kappa}{2j(j+1)}\sqrt{(j+j_3+1)(j-j_3)}\, Y_{j, j_3+1}(\polar,\phi)\\
        & -\frac{1}{2j}\sqrt{\frac{(j+\frac{\kappa}{2})(j-\frac{\kappa}{2})(j-j_3)
                        (j-j_3-1)}{(j-\frac{1}{2})(j+\frac{1}{2})}}\, Y_{j-1, j_3+1}(\polar, \phi).
\end{split}\raisetag{2\baselineskip}
\end{align}
\end{subequations}
Applying these identities to Eq.~\eqref{fermPosSpace}, one finds,
after some tedious algebra, that
\begin{subequations}\label{fermWavefcns}
\begin{align}
\begin{split}
\braket{r, \polar, \phi}{E; 1; j-\tfrac{1}{2}, j, j_3} &= \frac{1}{\sqrt{2mE}}\Biggl\{
        2\frac{\radwav'_- - \frac{\radwav_-}{r}(j-\tfrac{1}{2})}{2j+1}
                        \sqrt{(j+\tfrac{1}{2})^2-\tfrac{\kappa^2}{4}}
                        \fang{j+\frac{1}{2}}{j}{j_3}\\
        &\qquad         +\left(\kappa\frac{\radwav'_- - \frac{\radwav_-}{r}(j-\tfrac{1}{2})}{2j+1}
                -\param \radwav_-\right)\fang{j-\frac{1}{2}}{j}{j_3}
        \Biggr\},
\end{split}\raisetag{2\baselineskip}\\
\begin{split}
\braket{r,\polar, \phi}{E; 1;j+\tfrac{1}{2}, j, j_3} &= \frac{1}{\sqrt{2mE}}\Biggl\{
        2\frac{\radwav'_+ + \frac{\radwav_+}{r}(j+\tfrac{3}{2})}{2j+1}
                        \sqrt{(j+\tfrac{1}{2})^2-\tfrac{\kappa^2}{4}}
                        \fang{j-\frac{1}{2}}{j}{j_3}\\
        &\qquad -\left(\kappa\frac{\radwav'_+ + \frac{\radwav_+}{r}(j+\tfrac{3}{2})}{2j+1}
                        +\param \radwav_+\right)\fang{j+\frac{1}{2}}{j}{j_3}
        \Biggr\},
\end{split}\raisetag{2\baselineskip}
\end{align}
\end{subequations}
where recall that
\begin{equation}
\radwav_\pm = \radwav_{E, j\pm\frac{1}{2}}(r),
\end{equation}
the radial dependence of the appropriate bosonic wavefunction, and the
fermionic harmonics are given in Eq.~\eqref{Yjjm}.

\section{Unitary Representations of SO(3,1) and its Contraction}
\label{sec:SO31}

In this section we review the unitary representations of
SO(3,1) and of the Galilei group.  For more details, see
e.g.~\cite{gelfand}.

Consider the six dimensional algebra of {\em hermitian\/} operators,
\begin{align} \label{SO31}
\com{J^i}{J^j} &= i \epsilon^{ijk} J^k, &
\com{J^i}{K^j} &= i \epsilon^{ijk} K^k, &
\com{K^i}{K^j} &= -i \alpha \epsilon^{ijk} J^k, & 
i=1,2,3, &\quad \alpha \geq 0.
\end{align}
For $\alpha > 0$, $K$ can be rescaled to obtain the standard SO(3,1)
generators; $\alpha=0$ corresponds to the contracted algebra.
So without loss of generality, one could restrict to $\alpha=0,1$.

The Casimirs of the algebra~\eqref{SO31} are
\begin{align} \label{SO31Casimirs}
C_1 &= \alpha \vec{J}^2 - \vec{K}^2, &
C_2 &= \vec{J} \cdot \vec{K}.
\end{align}
Because of the subalgebra generated by $\vec{J}$,
a unitary \irrep is given by a sum of unitary SU(2) representations.
So states in the unitary \irrep are labelled
\begin{equation}
\ket{c_1,c_2;j,j_3}
\end{equation}
where $c_1$ and $c_2$ are the values of the Casimirs~\eqref{SO31Casimirs},
and, of course, $j(j+1)$ is the value of the SU(2) Casimir $\vec{J}^2$ on
the state,
$j_3=-j,-j+1,\dotsc,j-1,j$ is the eigenvalue of $J^3$ on the state,
and $j$ is half a nonnegative integer.

Because $\vec{K}$ is a spin one operator, it will shift the value of $j$
at most by one.  The $\com{K^i}{J^j}$ commutators---or Clebsch-Gordan
coefficients---yield
\begin{subequations} \label{KonSO31States}
\begin{align}
\begin{split}
K^3 \ket{c_1,c_2;j,j_3} &= 
-\sqrt{j^2-j_3^2} A^{c_1,c_2}_{j-1,j} \ket{c_1,c_2;j-1,j_3}
+ j_3 A^{c_1,c_2}_{j,j} \ket{c_1,c_2;j,j_3}
\\ & \qquad
+ \sqrt{(j+1)^2-j_3^2} A^{c_1,c_2}_{j+1,j} \ket{c_1,c_2;j+1,j_3}, 
\end{split}
\\
\begin{split}
K^+ \ket{c_1,c_2;j,j_3} &= 
-\sqrt{(j-j_3)(j-j_3-1)} A^{c_1,c_2}_{j-1,j} \ket{c_1,c_2;j-1,j_3+1}
\\ & \qquad
+ \sqrt{(j-j_3)(j+j_3+1)} A^{c_1,c_2}_{j,j} \ket{c_1,c_2;j,j_3+1}
\\ & \qquad
- \sqrt{(j+j_3+1)(j+j_3+2)} A^{c_1,c_2}_{j+1,j} \ket{c_1,c_2;j+1,j_3+1},
\end{split}
\\
\begin{split}
K^- \ket{c_1,c_2;j,j_3} &= 
\sqrt{(j+j_3)(j+j_3-1)} A^{c_1,c_2}_{j-1,j} \ket{c_1,c_2;j-1,j_3-1}
\\ & \qquad
+ \sqrt{(j+j_3)(j-j_3+1)} A^{c_1,c_2}_{j,j} \ket{c_1,c_2;j,j_3-1}
\\ & \qquad
+ \sqrt{(j-j_3+1)(j-j_3+2)} A^{c_1,c_2}_{j+1,j} \ket{c_1,c_2;j+1,j_3-1},
\end{split}
\end{align}
\end{subequations}
where the coefficients $A^{c_1,c_2}_{j',j}$ remain to be determined.
By appropriately choosing the relative phases of~$\ket{c_1,c_2;j,j_3}$
and using $(K^-)^\dagger = K^+$, one can take
\begin{align} \label{phasechoice}
A^{c_1,c_2}_{j,j} &\in \ZR, &
A^{c_1,c_2}_{j-1,j} &= -A^{c_1,c_2}_{j,j-1} \geq 0.
\end{align}
Because $j\geq 0$, there is a minimum value, $j_0$, of $j$.
Except for the trivial representation, $j=j_0=0$, there is no maximum value
of $j$ because unitary \irrep{s}
of noncompact algebras are infinite dimensional.
All other values
of $j$ in the \irrep are obtained by applying $\vec{K}$ one or more times,
and so typically $j=j_0,j_0+1,\dotsc$.  (The exceptions are the trivial
singlet representation and the representations with $\vec{K}=0$ if $\alpha=0$.)

Evaluating the Casimir $C_2$ on a state, using
Eq.~\eqref{KonSO31States} yields
\begin{equation} \label{c2}
A_{j,j}^{c_1,c_2} = \frac{c_2}{j(j+1)}.
\end{equation}
(In matrix elements, this will be multiplied by zero if $j=0$ and so
is effectively well-defined.)
Similarly, the Casimir $C_1$ gives the less trivial relation
\begin{equation} \label{c1}
\alpha j(j+1) - c_1 = -j(2j-1) A^{c_1,c_2}_{j-1,j} A^{c_1,c_2}_{j,j-1}
 + j(j+1) (A^{c_1,c_2}_{j,j})^2
 - (j+1)(2j+3) A^{c_1,c_2}_{j,j+1} A^{c_1,c_2}_{j+1,j}.
\end{equation}
The commutator $\com{K^+}{K^-} = -2 \alpha J^3$ translates to
\begin{equation} \label{[Kp,Km]}
-(2j-1) A^{c_1,c_2}_{j-1,j} A^{c_1,c_2}_{j,j-1}
+ (A^{c_1,c_2}_{j,j})^2
+ (2j+3) A^{c_1,c_2}_{j+1,j} A^{c_1,c_2}_{j,j+1} = -\alpha
  \text{ or } j=0.
\end{equation}
Eqs.~\eqref{c1}, \eqref{[Kp,Km]},~\eqref{c2}
and~\eqref{phasechoice} combine to yield
\begin{equation} \label{Ajj+1}
\bigl[A_{j,j+1}^{c_1,c_2}\bigr]^2 = 
\frac{\alpha j(j+1)^2(j+2) - (j+1)^2 c_1 - c_2^2}{(j+1)^2(2j+1)(2j+3)}.
\end{equation}

In order that $j=j_0$ be the smallest value of $j$ that appears in the
unitary \irrep, it is necessary that $A^{c_1,c_2}_{j_0-1,j_0} = 0$.
Conversely, $A_{j-1,j}=0$ when
\begin{equation}
\alpha j^2
= \tfrac{1}{2} (c_1+\alpha) \pm \sqrt{\tfrac{1}{4}(c_1+\alpha)^2 + c_2^2}.
\end{equation}
From this it is deduced that
\begin{itemize}
\item For $c_2\neq 0$,
only the plus sign gives a valid $j^2$, and so, as anticipated
above, all SU(2) representations $j=j_0,j_0+1,\dotsc$ appear in the
unitary \irrep.  Moreover,
\hbox{$\alpha j_0^2 
= \tfrac{1}{2} (c_1+\alpha) + \sqrt{\tfrac{1}{4}(c_1+\alpha)^2 + c_2^2} > 0$};
i.e. $j_0=0$ is incompatible with $c_2\neq 0$.
Also, $c_1= \alpha (j_0^2-1) - \frac{c_2^2}{j_0^2}$.
\item If $c_2=0$ then $A^{c_1,c_2}_{j-1,j} = 0$ for $j=0$ or 
$\alpha j^2=\alpha+c_1$.
Eq.~\eqref{Ajj+1} implies that $c_1\leq 0$ if $j_0=c_2=0$.
Thus,
\begin{itemize}
\item $c_1=0=c_2$ is the singlet representation $j=j_0=0$, unless $\alpha=0$.
If $\alpha=0$ then $c_1=0=c_2$ implies $j=j_0$, but $j_0$ may take
any value.  This corresponds to the partially trivial
unitary \irrep that is a unitary \irrep
of SU(2) with $\vec{K}=0$.
\item If $j_0=0$ and $c_1<0$ ($c_2=0$), there
is no {\em positive integer\/} value of $j$, except $j=j_0$,
for which $A^{c_1,c_2}_{j-1,j}=0$.  Thus, again $j=j_0,j_0+1,\dotsc$.
\item If $j_0>0$ and $c_2=0$, then $c_1=\alpha (j_0^2-1)$.
So if $\alpha=0$, then $c_2=0$ implies $c_1=0$, corresponding to the partially
trivial \irrep.
\end{itemize}
\end{itemize}

Finally, it is convenient to write
\begin{equation}
c_2 = j_0 \zeta, \qquad \zeta \in \ZR.
\end{equation}
When $j_0 \neq 0$, the above considerations imply
\begin{equation}
c_1 = \alpha (j_0^2-1) - \zeta^2.
\end{equation}
If $j_0=0$ (and therefore $c_2=0$),
and $\alpha \neq 0$, then $c_1$ is arbitrary.

To summarize, if the Casimir
\begin{equation}
C_2 = \vec{J}\cdot \vec{K} = j_0 \zeta \neq 0,
\end{equation}
then the representation consists of states with $j=j_0,j_0+1,\dotsc$,
and the other Casimir has value
\begin{equation}
C_1 = \alpha \vec{J}^2 - \vec{K}^2 = \alpha (j_0^2-1) - \zeta^2.
\end{equation}
This is the case needed in the main text; the other cases have
$\vec{J}\cdot\vec{K}=0$.

\section{Normalization of the Unbound Bosonic States} \label{sec:norm}

In this section the computation of the normalization of the
wavefunction~\eqref{bosonicNBWF} is given.
This is accomplished by computing the norm of the radial function
\begin{equation}
\radwav_{E,j}(r) = \frac{1}{r}
   M_{-i\frac{\kappa\param}{\en},j+\frac{1}{2}}(-i \en r),
\qquad
\en \equiv \sqrt{2mE-\param^2}.
\end{equation}

The first step in the calculation is the integral~\cite[Eq.~(7.622.3)]{gr},
\begin{subequations}
\begin{multline} \label{GRMint}
\int_0^\infty dx\, x^{\nu-1} e^{-bx}
  M_{\lambda_1,\mu_1-\frac{1}{2}}(a_1 x)
  M_{\lambda_2,\mu_2-\frac{1}{2}}(a_2 x)
= a_1^{\mu_1} a_2^{\mu_2} [b+\tfrac{1}{2}(a_1+a_2)]^{-\nu-\mu_1-\mu_2}
  \Gamma(\nu+\mu_1+\mu_2)
\\ \times
  F_2\bigl(\nu+\mu_1+\mu_2;\mu_1-\lambda_1,\mu_2-\lambda_2;2\mu_1,2\mu_2;
      \tfrac{a_1}{b+\frac{1}{2}(a_1+a_2)},\tfrac{a_2}{b+\frac{1}{2}(a_1+a_2)}
\bigr),\\
\re\left(\nu+\mu_1+\mu_2\right)>0,
\re\left(b\pm\tfrac{1}{2}a_1\pm\tfrac{1}{2}a_2\right) > 0,
\end{multline}
using Bailey's~\cite{bailey} notation for Appell's second
hypergeometric function of two variables,
\begin{equation} \label{defF2}
F_2(\alpha;\beta,\beta';\gamma,\gamma';x,y)
\equiv \frac{\Gamma(\gamma)\Gamma(\gamma')}{
\Gamma(\alpha)\Gamma(\beta)\Gamma(\beta')}
\sum_{m,n=0}^\infty \frac{\Gamma(\alpha+n+m)\Gamma(\beta+n)\Gamma(\beta'+m)}{
n!m!\Gamma(\gamma+n)\Gamma(\gamma'+m)}x^n y^m.
\end{equation}
\end{subequations}
The reader who wishes to derive the integral~\eqref{GRMint} can easily do
so by writing the left-hand side in terms of the confluent hypergeometric
series, and integrating term by term in $r$ to immediately obtain the
double series~\eqref{defF2}.
It will be possible to simplify the result by using the identity
\begin{equation} \label{normFinalId}
\begin{split}
F_2&(\alpha+1;\beta,\beta';\alpha,\alpha;x,y)
= \frac{1}{\alpha(\beta-\beta')} (1-x-y)^{\alpha-\beta-\beta'-1}
  (1-x)^{\beta'-\alpha+1} (1-y)^{\beta-\alpha+1}
\\ & \times
\left\{
\beta \frac{\alpha-2\beta'-x(\alpha-\beta-\beta')}{1-x}
  {_2}F_1\bigl(\alpha-\beta-1,\alpha-\beta';\alpha;\tfrac{xy}{(1-x)(1-y)}\bigr)
\right. \\ & \qquad \left.
+ \beta' \frac{-\alpha+2\beta+y(\alpha-\beta-\beta')}{1-y}
  {_2}F_1\bigl(\alpha-\beta,\alpha-\beta'-1;\alpha;\tfrac{xy}{(1-x)(1-y)}\bigr)
\right\}.
\end{split}
\end{equation}

The identity~\eqref{normFinalId} is derived from the identity~\cite{bailey},
\begin{equation} \label{appellfmla}
F_2(\alpha;\beta,\beta';\alpha,\alpha;x,y)
= (1-x)^{-\beta} (1-y)^{-\beta'}
{_2F_1}\bigl(\beta,\beta';\alpha;\tfrac{x y}{(1-x)(1-y)}\bigr),
\end{equation}
which can be confirmed by writing the right-hand side as a power series in $x$
and $y$.  In order to put the Appell function on the left-hand side
of~\eqref{normFinalId} into ones of the form of
that on the left-hand side of~\eqref{appellfmla},
write
\begin{multline}
\Gamma(\alpha+n+m+1)\Gamma(\beta+n)\Gamma(\beta'+m)
\\= \left[(\alpha-\beta-\beta') + (\beta+n) + (\beta'+m)\right]
\Gamma(\alpha+n+m)\Gamma(\beta+n)\Gamma(\beta'+m)
\\= (\alpha-\beta-\beta') \Gamma(\alpha+n+m)\Gamma(\beta+n)\Gamma(\beta'+m)
+ \Gamma(\alpha+n+m)\Gamma(\beta+1+n)\Gamma(\beta'+m)
\\+ \Gamma(\alpha+n+m)\Gamma(\beta+n)\Gamma(\beta'+1+m),
\end{multline}
which implies
\begin{multline} \label{shiftAppell}
F_2(\alpha+1;\beta,\beta';\alpha,\alpha;x,y)
= \frac{\alpha-\beta-\beta'}{\alpha} F_2(\alpha;\beta,\beta';\alpha,\alpha;x,y)
 + \frac{\beta}{\alpha} F_2(\alpha;\beta+1,\beta';\alpha,\alpha;x,y)
\\ + \frac{\beta'}{\alpha} F_2(\alpha;\beta,\beta'+1;\alpha,\alpha;x,y).
\end{multline}
The Appell functions on the right-hand side of~\eqref{shiftAppell} can then
be written in terms of ordinary hypergeometric functions via
Eq.~\eqref{appellfmla}.
In fact, the identity,
\begin{equation} \label{2F1shiftId}
{_2}F_1(\beta,\beta';\alpha;z) =
\frac{\beta}{\beta-\beta'} {_2}F_1(\beta+1,\beta';\alpha;z)
- \frac{\beta'}{\beta-\beta'} {_2}F_1(\beta,\beta'+1;\alpha;z),
\end{equation}
whose derivation is similar to that
of~\eqref{shiftAppell},
allows one to write the hypergeometric function from the first term
in Eq.~\eqref{shiftAppell} in terms of the other two.
If one also uses~\cite{bailey}%
\begin{equation} \label{2F1:reindex}
{_2}F_1(\beta,\beta';\alpha;z)
= (1-z)^{\alpha-\beta-\beta'}{_2F_1}(\alpha-\beta,\alpha-\beta';\alpha;z),
\qquad \re \gamma > \re \alpha,\re \beta > 0,
\end{equation}
then one finally obtains~\eqref{normFinalId}.

With these formulas,
and upon introducing a factor to ensure convergence of the integral, the inner
product of interest is
\begin{equation}
\begin{split}
N_{\en,\en',j} &= \lim_{\delta\rightarrow 0^+}\int_0^\infty dr r^2 e^{-\delta r}
  \radwav_{\en,j}(r) \radwav^*_{\en',j}(r)
= \lim_{\delta\rightarrow 0^+} \int_0^\infty dr e^{-\delta r}
  M_{-i \frac{\kappa\param}{\en},j+\frac{1}{2}}(-i \en r)
  M_{i \frac{\kappa\param}{\en'},j+\frac{1}{2}}(i \en' r)\\
&= \lim_{\delta\rightarrow 0^+}
\frac{\delta}{\delta^2+\frac{1}{4}(\en-\en')^2}
\frac{\left[\delta+\frac{i}{2}(\en-\en')\right]^{i
  \frac{\kappa\param}{\en\en'}(\en-\en')}
}{\delta-\frac{i}{2}(\en-\en')}
\frac{2 (2j+1)! (\en\en')^{j+1}}{\left[ \delta^2 +
    \frac{1}{4}(\en+\en')^2 \right]^j}
\frac{[\delta-\frac{i}{2}(\en+\en')]^{
  i\frac{\kappa\param}{\en}}}{
   [\delta+\frac{i}{2}(\en+\en')]^{
  i\frac{\kappa\param}{\en'}}}
\\ & \qquad \times
\left\{ \frac{\en}{\en+\en'}
  \frac{j+1+i\frac{\kappa\param}{\en}}{
        \delta+\frac{i}{2}(\en+\en')}
 {_2}F_1\bigl(j-i\tfrac{\kappa\param}{\en},j+1+i\tfrac{\kappa\param}{
    \en'};2j+2;\tfrac{\en\en'}{\delta^2+\frac{1}{4}(
      \en+\en')^2}\bigr)
\right. \\ & \qquad \qquad \left.
+ \frac{\en'}{\en+\en'}
  \frac{j+1-i\frac{\kappa\param}{\en'}}{
        \delta-\frac{i}{2}(\en+\en')}
 {_2}F_1\bigl(j+1-i\tfrac{\kappa\param}{\en},j+i\tfrac{\kappa\param}{
    \en'};2j+2;\tfrac{\en\en'}{\delta^2+\frac{1}{4}(
      \en+\en')^2}\bigr)
\right\}.
\end{split}
\end{equation}
To proceed, observe that the first factor appears to imply the expected
$\delta$-function, whereupon the hypergeometric functions are evaluated
at unit argument.  However, the denominator of the second factor
provides a complication, which, fortunately, is overcome because it
turns out that the leading-order contribution from the quantity
in curly brackets vanishes.  Thus, let us expand the quantity in curly
brackets by using%
\footnote{This is a consequence of the identity
$\frac{d{_2F_1}(\beta,\beta';\alpha;z)}{dz} = 
\frac{\beta\beta'}{\alpha}{_2F_1}(\beta+1,\beta'+1;\alpha+1;z)$,
which follows immediately from shifting the index of summation in
the hypergeometric series, and from Gauss' formula~\cite{bailey}
$_2F_1(\beta,\beta';\alpha;1)= \frac{\Gamma(\alpha)\Gamma(\alpha-\beta-\beta')}{
        \Gamma(\alpha-\beta)\Gamma(\alpha-\beta')}$%
%
.}
\begin{equation} \label{2F1:expand}
{_2F_1}(\beta,\beta';\alpha;1-z)
= \frac{\Gamma(\alpha)\Gamma(\alpha-\beta-\beta')}{
        \Gamma(\alpha-\beta)\Gamma(\alpha-\beta')}
\left[1 - \frac{\beta \beta'}{\alpha-\beta-\beta'-1} z + \order{z^2}\right],
\re (\alpha-\beta-\beta') > 1;
\end{equation}
we trust the reader will forgive us for applying this formula on the
(excluded) boundary of the regime of validity.
(The extra step~\eqref{2F1:reindex} ensured that the parameters were 
at least on the
boundary, and not completely outside the regime of validity
of~\eqref{2F1:expand}.)

Then,
\begin{multline}
N_{\en,\en',j} = 
\lim_{\delta\rightarrow 0^+}
\frac{\left[\delta+\frac{i}{2}(\en-\en')\right]^{i
  \frac{\kappa\param}{\en\en'}(\en-\en')}
}{\left[\delta^2 + \frac{1}{4}(\en+\en')^2\right]^{j+1}}
\frac{[\delta-\frac{i}{2}(\en+\en')]^{
  i\frac{\kappa\param}{\en}}}{
   [\delta+\frac{i}{2}(\en+\en')]^{
  i\frac{\kappa\param}{\en'}}}
\frac{2 (2j+1)!^2 (\en\en')^{j+1}
\Gamma\bigl(1-i\frac{\kappa\param}{\en\en'}(\en-\en')\bigr)}{
  \Gamma(j+1+i\frac{\kappa\param}{\en})\Gamma(j+1-i\frac{\kappa\param}{\en'})}
\\ \times
\Biggl\{
\frac{\delta}{\delta^2+\frac{1}{4}(\en-\en')^2}
- \frac{\delta}{\delta^2+\frac{1}{4}(\en+\en')^2}
    \frac{(j-i\frac{\kappa\param}{\en})(j+i\frac{\kappa\param}{\en'})+j}{
    -i\frac{\kappa\param}{\en\en'}(\en-\en')}
\\
+ \frac{\delta^2 + \frac{i}{2}\delta (\en-\en')}{
\delta^2+\frac{1}{4}(\en-\en')^2}
\frac{\delta}{\delta^2+\frac{1}{4}(\en+\en')^2}
    \frac{\en\en'}{-i(\en-\en')}
+ \order{\delta\bigl[\delta + \frac{i}{2}(\en-\en')\bigr]}
\Biggr\}.
\end{multline}
By definition, $\en,\en'>0$, so the denominator of the first factor is regular
as $\delta\rightarrow 0^+$.  It is the objects in the curly brackets that
are interesting in the limit.  The first term is a $\delta$-function.
The second term vanishes in the limit.  Splitting up the third term
according to the numerator in the first factor, the $\delta^2$ term vanishes
in the limit, and in the second term, the factor of $(\en-\en')$ cancels
between numerator and the (last) denominator, leaving a $\delta$-function
(and a factor of $-\frac{1}{2}$.)  The neglected terms clearly vanish
in the limit.

Since
\begin{equation}
\lim_{\delta\rightarrow 0^+}
\frac{[\delta-\frac{i}{2}(\en+\en')]^{
  i\frac{\kappa\param}{\en}}}{
   [\delta+\frac{i}{2}(\en+\en')]^{
  i\frac{\kappa\param}{\en'}}}
= \frac{e^{-\frac{\pi i}{2}\frac{i\kappa \param}{\en}}}{
e^{\frac{\pi i}{2}\frac{i\kappa\param}{\en'}}}
= e^{\pi \frac{\kappa\param}{2\en\en'}(\en+\en')},
\end{equation}
we have finally found
\begin{equation}
N_{\en,\en',j} =
\pi\frac{(2j+1)!^2}{\abs{\Gamma(j+1+i\frac{\kappa\param}{\en})}^2}
e^{\pi \frac{\kappa\param}{\en}}
\delta(\en-\en'), \qquad
\delta(\en-\en') = \frac{\sqrt{2 m E - \param^2}}{m E}
\delta(E-E').
\end{equation}

\end{document}